\documentclass[11pt]{article}
\usepackage{jheppub}
\usepackage{subcaption}
\usepackage{tikz}

\DeclareMathOperator{\Tr}{Tr}

\newcommand{\ri}{{\rm i}}

\def\th{\theta}

\def\ep{\epsilon}

\newcommand{\hf}{\frac{1}{2}}
\newcommand{\qu}{\frac{1}{4}}

\def\til#1{\widetilde{#1}}
\def\si{\sigma}

\def\del{\partial}

\def\lap{\Delta}

\def\la{\lambda}

\def\bt{\beta}

\def\al{\alpha}

\newcommand{\Om}{\Omega}
\def\rt#1{\sqrt{#1}}

\begin{document}

\title{Phase diagram of $q$-deformed Yang-Mills theory on $S^2$ at non-zero $\theta$-angle}

\author{Kazumi Okuyama}

\affiliation{Department of Physics, 
Shinshu University, Matsumoto 390-8621, Japan}

\emailAdd{kazumi@azusa.shinshu-u.ac.jp}

\abstract{
We study the phase diagram of $q$-deformed Yang-Mills theory
on $S^2$ at non-zero $\theta$-angle using the exact partition function at finite $N$.
By evaluating the exact partition function numerically,
we find evidence for 
the existence of a series of phase transitions at non-zero
$\theta$-angle as conjectured in [hep-th/0509004]. 
}

\maketitle

\renewcommand{\thefootnote}{\arabic{footnote}}
\setcounter{footnote}{0}
\setcounter{section}{0}

\section{Introduction}\label{sec:intro}
The topological $\th$-angle
plays an important role in the dynamics of gauge theories and $\si$-models
in various dimensions (see e.g. \cite{Callan:1976je,Coleman:1976uz,DAdda:1978vbw,Witten:1993yc,Gaiotto:2017yup} and references therein).
More than a decade ago, the phase diagram of $q$-deformed Yang-Mills theory ($q$YM) on $S^2$
at non-zero $\th$-angle was conjectured in \cite{Jafferis:2005jd}.
In this paper, we will study the phase diagram of $q$YM on $S^2$
using the exact partition function at finite $N$.
In general, 
exact result is a very powerful tool to analyze the non-perturbative aspects of gauge theories,
which are usually beyond reach by other means.
This approach was successfully applied to many examples, including
the ABJM theory on $S^3$ \cite{Hatsuda:2012hm,Hatsuda:2012dt,Hatsuda:2013gj,Hatsuda:2013oxa}
and the Gross-Witten-Wadia (GWW) unitary matrix model \cite{Marino:2008ya,Okuyama:2017pil,Ahmed:2017lhl}.

In \cite{Jafferis:2005jd,Arsiwalla:2005jb,Caporaso:2005ta},
it was found that $q$YM on $S^2$ in the large $N$ 't Hooft limit
has a third order phase transition at some critical value
$t=t_c$ of the 't Hooft coupling 
when $\th=0$. The physical mechanism of this phase transition is 
essentially the same as that of ordinary (undeformed) Yang-Mills 
theory on $S^2$
found by Douglas and Kazakov \cite{Douglas:1993iia}.
As shown in  \cite{Jafferis:2005jd,Arsiwalla:2005jb,Caporaso:2005ta}, 
at the critical point
the instanton contribution becomes comparable with the zero-instanton sector
and the phase transition is triggered by the instanton condensation,
in much the same way as the undeformed Yang-Mills theory \cite{Gross:1994mr}.

In \cite{Jafferis:2005jd}, it was further conjectured that
the phase diagram of $q$YM at non-zero $\th$ has an intricate structure
(see Figure 3 in \cite{Jafferis:2005jd}):
there is a series of phase transition curves on the $t\text{-}\th$ plane
which accumulate at the point $(t,\th)=(0,\pi)$.
Each transition curve corresponds to the exchange of dominance of different 
instanton sectors.
In this paper we will examine this conjecture using the exact partition function of $q$YM
at finite $N$.
By evaluating the exact partition function numerically, we find
evidence for the conjectured phase diagram in \cite{Jafferis:2005jd}.
We find that 
the {\it prefactor} of instanton contribution is
important for the understanding of the phase structure at non-zero $\th$.
However, it turns out that it is difficult to find the analytic form
of this prefactor in the $q$-deformed case.
Instead, in section \ref{sec:undeform} we consider the undeformed case 
where the prefactor of 1-instanton
contribution is already known \cite{Gross:1994mr}, and we 
propose an analytic form of the first two transition curves at non-zero $\th$ 
and check this proposal numerically.

This paper is organized as follows.
In section \ref{sec:exact-z}, we write down the 
exact partition function of $q$YM on $S^2$ as a determinant of
$N\times N$ matrix.
Using this exact result, we 
study the behavior of free energy at $\th=0$ and
confirm
the phase transition found in \cite{Jafferis:2005jd,Arsiwalla:2005jb,Caporaso:2005ta}.
In section \ref{sec:inst}, we study the instanton contribution
using our exact result at finite $N$.
In section \ref{sec:phase}, we numerically compute the free energy at non-zero $\th$
and find evidence that indeed there is a series of phase transitions
coming from the exchange of dominance of different instanton sectors, as conjectured in
\cite{Jafferis:2005jd}.
 In section \ref{sec:undeform}, we consider the phase diagram of undeformed theory at
$\th\ne0$ 
and propose an analytic form of the first two transition curves. 
Finally, we conclude in section \ref{sec:discussion}
with some discussion of the future directions.
\section{Exact partition function at finite $N$}\label{sec:exact-z}
In this section, we consider the exact partition function of $q$YM on $S^2$
at finite $N$.
This section is mostly a review of the known results.

The $q$-deformed $U(N)$ Yang-Mills theory on 
$S^2$ naturally appears as a worldvolume theory on $N$ D4-branes
on a non-compact Calabi-Yau $X_p$
\begin{align}
 X_p:~\mathcal{O}(-p)\oplus  \mathcal{O}(p-2)\to \mathbb{P}^1,
\end{align}
where the D4-branes in question are wrapping 
around the base $\mathbb{P}^1$ and one of the fiber
$\mathcal{O}(-p)$ of $X_p$ \cite{Aganagic:2004js}. 
It is argued in \cite{Vafa:2004qa,Aganagic:2004js}
that  the path integral of
D4-brane worldvolume theory localizes to the 
$q$-deformed Yang-Mills theory on the base $\mathbb{P}^1=S^2$.
This partition function of D4-brane theory
is identified as the partition function of 4d black holes made of the bound states of
D4, D2, D0-branes, 
which in turn is 
related to the topological string partition function on
$X_p$ via the OSV conjecture \cite{Ooguri:2004zv}.

It is known that $q$YM on any Riemann surface can be solved exactly \cite{Aganagic:2004js}
in a similar manner as the ordinary undeformed 2d Yang-Mills theory \cite{Migdal:1975zg}.
The partition function of $U(N)$
$q$YM on $S^2$ is given by
\begin{align}
 Z_N=
\frac{1}{N!}\sum_{n_i\in\mathbb{Z}+\frac{\ep}{2}}\prod_{i<j}[n_i-n_j]^2
\exp\left(-\frac{g_sp}{2}\sum_{i=1}^Nn_i^2+\ri\th\sum_{i=1}^Nn_i\right), 
\label{eq:z-sum}
\end{align}
where $[n]$ denotes the $q$-integer
\begin{align}
 [n]=\mathbf{q}^{\frac{n}{2}}-\mathbf{q}^{-\frac{n}{2}},\qquad
\mathbf{q}=e^{-g_s},
\end{align}
and $\ep$ is given by
\begin{align}
 \ep=\Biggl\{
\begin{aligned}
 &0,\quad (\text{odd}~N),\\
&1,\quad (\text{even}~N).\\
\end{aligned}
\end{align}
In other words, the summation of $n_i$ in \eqref{eq:z-sum} runs over 
integers for odd $N$ and half-integers for even $N$.
\footnote{Our $n_i$ is related to the rows $\{\la_1,\cdots,\la_N\}$ of Young diagram  by
\begin{align}
 n_i=\la_i-i+\frac{N+1}{2}.
\end{align}
It follows that $n_i\in\mathbb{Z}+\frac{\ep}{2}$ since $\la_i$ is integer.
}
The overall normalization of
the partition function is ambiguous;
in \eqref{eq:z-sum}
we followed the convention of \cite{Jafferis:2005jd}.
As we will show below, we can rewrite this partition function as a
determinant of $N\times N$ matrix. To see this, we first notice that 
the factor $\prod_{i<j}[n_i-n_j]$ in \eqref{eq:z-sum}
is basically the Vandermonde determinant and it is rewritten as 
\begin{align}
 \prod_{i<j}[n_i-n_j]=
\sum_{\si\in S_N}(-1)^\si \prod_{i=1}^N \mathbf{q}^{(i-\frac{N+1}{2})n_{\si(i)}}.
\label{eq:vander}
\end{align}
This relation can also be understood as the Weyl denominator formula of $U(N)$ gauge group.
Squaring the above expression \eqref{eq:vander}
we get a sum over two permutations,
but one of them can be trivialized by
using the invariance of $\sum_i n_i^2$ and $\sum_i n_i$ under the permutation of $n_i$.
In this way \eqref{eq:z-sum} is rewritten as
\begin{align}
\begin{aligned}
 Z_N&=\sum_{n_i\in\mathbb{Z}+\frac{\ep}{2}}\sum_{\si\in S_N}
(-1)^\si \prod_{i=1}^N \mathbf{q}^{(i+\si(i)-N-1)n_{i}}e^{-\hf g_sp n_i^2+\ri\th n_i}\\
&=\sum_{\si\in S_N}
(-1)^\si \prod_{i=1}^N 
\vartheta_{3-\ep}\Bigl(\frac{\th+\ri g_s(i+\si(i)-N-1)}{2\pi},\frac{\ri g_sp}{2\pi}\Bigr),
\end{aligned}
\label{eq:ZN-SNsum}
\end{align} 
where the Jacobi theta function is defined by
\begin{align}
 \vartheta_{3-\ep}(v,\tau)=\sum_{n\in\mathbb{Z}+\frac{\ep}{2}}e^{2\pi\ri nv+\pi\ri n^2\tau}.
\end{align}
Finally, the sum over
$S_N$ in \eqref{eq:ZN-SNsum} reduces to the determinant
\begin{align}
 Z_N=
\det\Biggl[\vartheta_{3-\ep}\Bigl(\frac{\th+\ri g_s(i+j-N-1)}{2\pi},\frac{\ri g_sp}{2\pi}\Bigr)\Biggr]_{i,j=1,\cdots,N}.
\label{eq:ZN-det}
\end{align}
This is the main result of this section.\footnote{As far as we know, this expression has not appeared in the literature before.}
This determinant form of $Z_N$ 
is reminiscent of the exact partition function of GWW model \cite{Gross:1980he,Wadia:2012fr}. 
However, there is an important difference: 
the exact partition function of $q$YM in \eqref{eq:ZN-det} is given by a Hankel determinant $\det(a_{i+j})$, while
the exact partition function of GWW model is given by a Toeplitz determinant
$\det(b_{i-j})$.

We are interested in the behavior of this partition function 
in the large $N$ 't Hooft limit
\begin{align}
 g_s\to0,~~ N\to\infty,\quad t=g_sN:\text{fixed},
\end{align}
and we would like to study the genus expansion of free energy 
\begin{align}
 F=\log Z_N=\sum_{g=0}^\infty g_s^{2g-2}F_g(t).
\end{align}
In \cite{Jafferis:2005jd,Arsiwalla:2005jb,Caporaso:2005ta}, it is found 
that when $\th=0$ there is a third order phase transition
at the critical value $t=t_c$ of the 't Hooft coupling, where $t_c$ is given by
\begin{align}
 t_c=-2p\log\Bigl(\cos\frac{\pi}{p}\Bigr).
\label{eq:tc}
\end{align}
This phase transition occurs only for $p>2$
\cite{Jafferis:2005jd,Arsiwalla:2005jb,Caporaso:2005ta}, 
and we will assume
$p>2$ throughout this paper. 
In the rest of this section, we will consider the
behavior of free energy above ($t>t_c$) and below ($t<t_c$)
the phase transition using the exact partition function $Z_N$ at finite $N$.

\subsection{Strong coupling phase}
Let us first consider the strong coupling phase ($t>t_c$).
In the large $N$ limit, the eigenvalue distribution in
this phase is described by a two-cut solution of a certain matrix model
and the explicit form of the resolvent was constructed in \cite{Jafferis:2005jd,Arsiwalla:2005jb,Caporaso:2005ta}. However, it is not so straightforward to compute the genus-zero free energy
$F_0(t)$ from this solution.

To study this phase, it is 
convenient to regard the partition function \eqref{eq:z-sum} as a sum over configurations of $N$
non-relativistic fermions
\begin{align}
 Z_N=\sum_{n_1<\cdots<n_N}Z_{\vec{n}}=\sum_{n_1<\cdots<n_N}\prod_{i<j}[n_i-n_j]^2e^{-g_spE+\ri\th P}
\label{eq:Z-fermion}
\end{align}
where $\vec{n}=(n_1,\cdots,n_N)$ specifies the momentum of $N$ fermions 
and the total energy $E$ and the total momentum $P$ of fermions are
given by
\begin{align}
 E=\sum_{i=1}^N \hf n_i^2,\qquad P=\sum_{i=1}^N n_i.
\end{align}
These $N$ fermions are interacting through the factor $\prod_{i<j}[n_i-n_j]^2$. 

\begin{figure}[t]
\centering
\hskip10mm
\begin{tikzpicture}
\draw (-0.5,6)--(10.5,6);
\fill [gray!40] (0,6) circle [radius=2mm];
\fill [black] (1,6) circle [radius=2mm];
\fill [black] (2,6) circle [radius=2mm];
\fill [black] (3,6) circle [radius=2mm];
\fill [black] (4,6) circle [radius=2mm];
\fill [black] (5,6) circle [radius=2mm];
\fill [black] (6,6) circle [radius=2mm];
\fill [black] (7,6) circle [radius=2mm];
\fill [black] (8,6) circle [radius=2mm];
\fill [gray!40] (9,6) circle [radius=2mm];
\fill [gray!40] (10,6) circle [radius=2mm];
\coordinate (a1) at (1,5.8) node at (a1) [below] {$-n_F$};
\coordinate (a2) at (9.3,5.8) node at (a2) [below] {$n_F+1$};
\coordinate (a3) at (8,5.7) node at (a3) [below] {$n_F$};
\end{tikzpicture}
\caption{Maya diagram for the ground state.
The black nodes ($-n_F\leq n\leq n_F$) are occupied 
by fermions while the gray nodes ($|n|>n_F$) are empty.} 
\label{fig:gnd}
\end{figure}
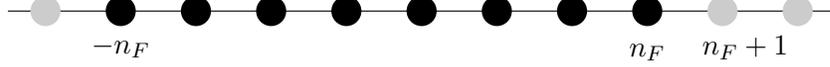

In the strong coupling phase,
we can compute \eqref{eq:Z-fermion} by summing over the fermion configurations 
from the small total energy $E$.
A similar computation has been performed for the undeformed Yang-Mills case in \cite{Douglas:1993iia}.
One can easily see that the ground state (lowest energy configuration) is given by 
\begin{align}
 \vec{n}_0=(-n_F,-n_F+1,\cdots, n_F-1,n_F),\qquad n_F=\frac{N-1}{2}.
\end{align}
The energy $E_0$ and the momentum $P_0$ of the ground state are given by
\begin{align}
 E_{0}=\hf\vec{n}_0^2=\frac{N^3-N}{24},\qquad P_0=0,
\end{align}
and the contribution of this ground state is
\begin{align}
 Z_{\text{gnd}}=Z_{\vec{n}_0}=e^{-g_sp E_0}\prod_{i<j}[i-j]^2=
e^{-\frac{g_sp}{24}(N^3-N)}\prod_{j=1}^{N-1}\Bigl(2\sinh\frac{g_sj}{2}\Bigr)^{2(N-j)}.
\label{eq:Zgnd}
\end{align}
One can visualize the configuration of 
fermions by the so-called Maya diagram as shown in Fig.~\ref{fig:gnd}:
the black nodes are occupied by the $N$ fermions and the gray nodes are empty.
In the ground state the nodes between $n=-n_F$ and $n=n_F$ are occupied;
the nodes at $n=\pm n_F$ can be thought of as the Fermi levels. 

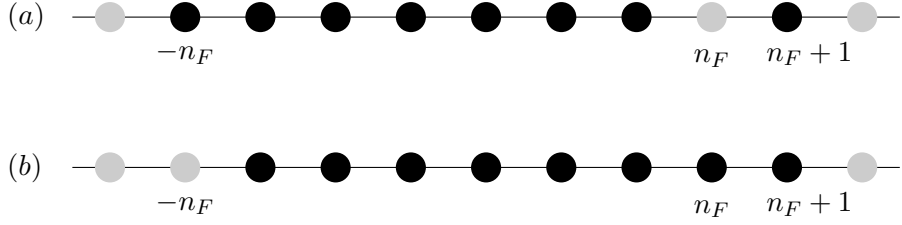
\begin{figure}[t]
\centering
\begin{tikzpicture}
\draw (-0.5,4)--(10.5,4);
\fill [gray!40] (1,4) circle [radius=2mm];
\fill [black] (2,4) circle [radius=2mm];
\fill [black] (3,4) circle [radius=2mm];
\fill [black] (4,4) circle [radius=2mm];
\fill [black] (5,4) circle [radius=2mm];
\fill [black] (6,4) circle [radius=2mm];
\fill [black] (7,4) circle [radius=2mm];
\fill [black] (8,4) circle [radius=2mm];
\fill [black] (9,4) circle [radius=2mm];
\fill [gray!40] (0,4) circle [radius=2mm];
\fill [gray!40] (10,4) circle [radius=2mm];
\coordinate (b1) at (1,3.8) node at (b1) [below] {$-n_F$};
\coordinate (b2) at (9.3,3.8) node at (b2) [below] {$n_F+1$};
\coordinate (b3) at (8,3.7) node at (b3) [below] {$n_F$};
\coordinate (B) at (-1.5,4) node at (B) [right] {$(b)$}; 
\draw (-0.5,6)--(10.5,6);
\fill [gray!40] (0,6) circle [radius=2mm];
\fill [black] (1,6) circle [radius=2mm];
\fill [black] (2,6) circle [radius=2mm];
\fill [black] (3,6) circle [radius=2mm];
\fill [black] (4,6) circle [radius=2mm];
\fill [black] (5,6) circle [radius=2mm];
\fill [black] (6,6) circle [radius=2mm];
\fill [black] (7,6) circle [radius=2mm];
\fill [gray!40] (8,6) circle [radius=2mm];
\fill [black] (9,6) circle [radius=2mm];
\fill [gray!40] (10,6) circle [radius=2mm];
\coordinate (a1) at (1,5.8) node at (a1) [below] {$-n_F$};
\coordinate (a2) at (9.3,5.8) node at (a2) [below] {$n_F+1$};
\coordinate (a3) at (8,5.7) node at (a3) [below] {$n_F$};
\coordinate (A) at (-1.5,6) node at (A) [right] {$(a)$};
\end{tikzpicture}
\caption{Examples of excited states with $\lap E=N/2$: (a) chiral,
(b) non-chiral. } 
\label{fig:excitation}
\end{figure}

We can also draw the Maya diagram for excited states as in Fig.~\ref{fig:excitation}.
The excitation energy $\lap E=E-E_0$ and the total momentum
of the states $(a)$ and $(b)$  in Fig.~\ref{fig:excitation} 
can be easily computed as
\begin{align}
 \begin{aligned}
  (a):&~\lap E=\hf (n_F+1)^2-\hf n_F^2=\frac{N}{2},\qquad
&P&=n_F+1-n_F=1,\\
(b):&~\lap E=\hf (n_F+1)^2-\hf (-n_F)^2=\frac{N}{2},\qquad
&P&=n_F+1-(-n_F)=N,
 \end{aligned}
\end{align}
and their contributions to the partition function are given by
\begin{align}
  \frac{Z_{(a)}}{Z_{\text{gnd}}}&=\frac{[N]^2}{[1]^2}
e^{-\frac{N}{2}g_sp+\ri\th},\qquad\frac{Z_{(b)}}{Z_{\text{gnd}}}=
e^{-\frac{N}{2}g_sp+\ri N\th}.
\end{align}
There are two more states with the same excitation energy $\lap E=N/2$
which are obtained by flipping the sign of momenta $n_i\to-n_i$ in Fig.~\ref{fig:excitation}.
In this way, we can compute
 the large $t$ expansion of partition function systematically
as
\begin{align}
\begin{aligned}
 &\frac{Z_N}{Z_{\text{gnd}}}=1+
\Biggl[2\cos\th\frac{[N]^2}{[1]^2}+2\cos N\th\Biggr]e^{-\frac{tp}{2}}\\
&+\Biggl[2\cos2\th \frac{\mathbf{q}^p[N]^2[N+1]^2+\mathbf{q}^{-p}[N]^2[N-1]^2}{[1]^2[2]^2}+\frac{[N-1]^2[N+1]^2}{[1]^4}
\\
&\quad+\Bigl(2\mathbf{q}^p\cos (N+1)\th +2\mathbf{q}^{-p}\cos(N-1)\th\Bigr)\frac{[N]^2}{[1]^2}\Biggr]e^{-tp}\\
&+\Biggl[2\cos3\th\Biggl(\frac{\mathbf{q}^{3p}[N]^2[N+1]^2[N+2]^2+\mathbf{q}^{-3p}[N]^2[N-1]^2[N-2]^2}{[1]^2[2]^2[3]^2}
+\frac{[N+1]^2[N]^2[N-1]^2}{[1]^4[3]^2}\Biggr)\\
&~~+2\cos\th\frac{\mathbf{q}^p[N+2]^2[N]^2[N-1]^2+\mathbf{q}^{-p}[N-2]^2[N]^2[N+1]^2}{[1]^4[2]^2}\\
&~~+2\cos(N+2)\th \frac{(\mathbf{q}^{3p}+\mathbf{q}^{-p})[N]^2[N+1]^2}{[1]^2[2]^2}+2\cos(N-2)\th \frac{(\mathbf{q}^{-3p}+\mathbf{q}^{p})[N]^2[N-1]^2}{[1]^2[2]^2}\\
&~~+2\cos N\th\frac{[N-1]^2[N+1]^2}{[1]^4}\Biggr]e^{-\frac{3tp}{2}} +\mathcal{O}(e^{-2tp}).
\end{aligned}
\label{eq:ZN-strong}
\end{align}

As a consistency check of our result \eqref{eq:ZN-strong},
we can take the limit 
\begin{align}
p\to\infty,\quad g=g_sp, ~A=gN:\text{fixed},
\label{eq:undeform-coupling}
\end{align}
in which the partition function of $q$YM reduces to 
the partition function of undeformed Yang-Mills theory \cite{Jafferis:2005jd,Arsiwalla:2005jb,Caporaso:2005ta}.
After taking this limit,  the free energy becomes
\begin{align}
 \begin{aligned}
  \log\frac{Z_N}{Z_{\text{gnd}}}&=\Bigl(1+\frac{A^2}{g^2}\Bigr)
\Bigl[2e^{-\frac{A}{2}}-e^{-A}+\frac{8}{3}e^{-\frac{3A}{2}}\Bigr]\\
&+\Biggl[-\frac{2A^3}{g^3}\sinh g+\Bigl(5+\frac{A^2}{g^2}\Bigr)\frac{2A^2}{g^2}\sinh^2\frac{g}{2}\Biggr]e^{-A}\\
&+\Biggl[-\Bigl(5+\frac{A^2}{g^2}\Bigr)\frac{8A^3}{3g^3}\sinh^3 g+\Bigl(13\cosh2g+26\cosh g-3\Bigr)
\frac{4A^2}{9g^2}\sinh^2\frac{g}{2}\\
&+\Bigl(11\cosh2g+22\cosh g-15\Bigr)
\frac{8A^4}{9g^4}\sinh^2\frac{g}{2}+\Bigl(\cosh g+2\Bigr)\frac{16A^6}{9g^6}\sinh^4\frac{g}{2}\Biggr]e^{-\frac{3A}{2}}+\mathcal{O}(e^{-2A}).
 \end{aligned}
\end{align}
Here we have set $\th=0$ for simplicity.
This can be further expanded in  the coupling $g$ as
\begin{align}
 \log\frac{Z_N}{Z_{\text{gnd}}}=\sum_{h=0}^\infty g^{2h-2}F_h(A),
\end{align}
where the first three terms read
\begin{align}
\begin{aligned}
F_0(A)&=2A^2e^{-\frac{A}{2}}+\Biggl(-1-2A+\frac{A^2}{2}\Biggr)A^2e^{-A}+\Biggl(
\frac{8}{3}+4 A^2-\frac{8 A^3}{3}+\frac{A^4}{3}\Biggr)A^2e^{-\frac{3A}{2}}+\mathcal{O}(e^{-2A}), \\
F_1(A)&= 2e^{-\frac{A}{2}}+\Biggl(-1+\frac{5A^2}{2}-\frac{A^3}{3}+\frac{A^4}{24}\Biggr)e^{-A}+
\Biggl(\frac{8}{3}+4 A^2
-\frac{40A^3}{3}+\frac{23 A^4}{3}-\frac{4 A^5}{3}+\frac{A^6}{9}\Biggr)e^{-\frac{3A}{2}}+\mathcal{O}(e^{-2A}),\\
F_2(A)&=\Biggl(\frac{5}{24}-\frac{A}{60}+\frac{A^2}{720}\Biggr)A^2e^{-A}+
\Biggl(\frac{14 }{3}-\frac{20 A}{3}
+\frac{221A^2}{90}-\frac{13 A^3}{45}+\frac{13 A^4}{720}\Biggr)A^2e^{-\frac{3A}{2}}
+\mathcal{O}(e^{-2A}).
\end{aligned}
\end{align}
As expected, the above $F_0(A)$ 
agrees with the genus-zero free energy of undeformed Yang-Mills theory
computed in \cite{Douglas:1993iia}\footnote{Our $F_0(A)$ and 
that in \cite{Douglas:1993iia} differ by a factor of $A^2$ which comes from the different
definition of the genus expansion. In \cite{Douglas:1993iia} 
the genus expansion is defined as the $1/N$ expansion, 
$\log Z=\sum_{h\geq0} N^{2-2h}F_h(A)$, 
while we expand the free energy in terms of $g=A/N$.
}.

\paragraph{Chiral partition function.} 
One can naturally distinguish the excitations as ``chiral'', ``non-chiral'', and ``anti-chiral'', 
as follows.
In the chiral excitation, changes from the ground state 
are allowed only near the positive Fermi level  $n=+n_F$.
In other words, a chiral excitation is a configuration
where the modes near the negative Fermi level are the same as the ground state
\begin{align}
 \vec{n}=(-n_F,-n_F+1, \cdots,n_{N-1},n_{N}).
\end{align}
Fig.~\ref{fig:excitation}(a) is an example of chiral excitation.
The anti-chiral excitation is the momentum flip $n_i\to-n_i$ of chiral excitation, i.e.
only the excitations near the negative Fermi level $n=-n_F$  are allowed.
If the excitation involves both of the Fermi levels $n=\pm n_F$ it is called non-chiral
(see Fig.~\ref{fig:excitation}(b) for an example of non-chiral excitation). This type of decomposition
was first considered in the undeformed Yang-Mills theory 
in \cite{Gross:1992tu,Gross:1993hu,Gross:1993yt}.

Now we can define the chiral partition function $Z_N^{+}$
by summing over the chiral excitations only.
$Z_N^{+}$ is easily found 
from the full partition function \eqref{eq:ZN-strong} 
by dropping the non-chiral and anti-chiral terms
\begin{align}
 \begin{aligned}
  \frac{Z_N^{+}}{Z_{\text{gnd}}}&=1+\frac{[N]^2}{[1]^2}e^{-\frac{Ng_sp}{2}+\ri\th}
+\frac{\mathbf{q}^p[N]^2[N+1]^2+\mathbf{q}^{-p}[N]^2[N-1]^2}{[1]^2[2]^2}e^{-Ng_sp+2\ri\th}\\
&+\Biggl(\frac{\mathbf{q}^{3p}[N]^2[N+1]^2[N+2]^2+\mathbf{q}^{-3p}[N]^2[N-1]^2[N-2]^2}{[1]^2[2]^2[3]^2}
+\frac{[N+1]^2[N]^2[N-1]^2}{[1]^4[3]^2}\Biggr)e^{-\frac{3Ng_sp}{2}+3\ri\th}\\
&+\mathcal{O}(e^{-2Ng_sp+4\ri\th}).
 \end{aligned}
\label{eq:Z+}
\end{align}  
It turns out that
the chiral free energy $\log Z^{+}_N$ can be organized into 
a double expansion in terms of  $Q$ and $\til{Q}$
in the 't Hooft limit, where $Q$ and $\til{Q}$ are defined by
\begin{align}
\begin{aligned}
Q&=e^{-T},~~ &T&=\frac{p-2}{2} Ng_s-\ri\th,\\
\til{Q}&=e^{-t},~~ &t&=Ng_s.
\end{aligned}
\end{align}
The terms involving $\til{Q}$ correspond to open string amplitudes
due to the additional D-brane insertions on $X_p$ \cite{Aganagic:2004js}.
The pure closed string amplitude is obtained by
discarding the $\til{Q}$ dependent terms from \eqref{eq:Z+}
\begin{align}
 \begin{aligned}
  \frac{Z_N^{+,\text{closed}}}{Z_{\text{gnd}}}&=1+\frac{1}{[1]^2}Q
+\frac{\mathbf{q}^{p-1}+\mathbf{q}^{-p+1}}{[1]^2[2]^2}Q^2
+\Biggl(\frac{\mathbf{q}^{3p-3}+\mathbf{q}^{-3p+3}}{[1]^2[2]^2[3]^2}
+\frac{1}{[3]^2[1]^4}\Biggr)Q^3+\mathcal{O}(Q^4),
 \end{aligned}
\end{align} 
and the closed string free energy $F^{\text{closed}}=\log(Z_N^{+,\text{closed}}/Z_{\text{gnd}})$ is
given by
\begin{align}
 F^{\text{closed}}=\sum_{n=1}^3\frac{Q^n}{n[n]^2}+\frac{[p][p-2]}{[1]^2[2]^2}Q^2
+\Bigl([p-1]^2+[1]^2+6\Bigr)\frac{[p][p-1]^2[p-2]}{[1]^2[2]^2[3]^2}Q^3+\mathcal{O}(Q^4).
\label{eq:chiral-F}
\end{align} 
One can show that the genus expansion of $F^{\text{closed}}$ in \eqref{eq:chiral-F}
reproduces the result of topological string 
on $X_p$  
\cite{Caporaso:2005fp,Caporaso:2006gk,Forbes:2006ab}.

\subsection{Weak coupling phase}\label{sec:weak}
Next consider the weak coupling phase ($t<t_c$).
To study the weak coupling phase, we should perform the modular $S$-transformation
of the Jacobi theta function in \eqref{eq:ZN-det}.
Using the formula
\begin{align}
 \vartheta_{3-\ep}(v,\tau)=(-\ri\tau)^{-\hf}e^{-\frac{\pi\ri v^2}{\tau}}
\vartheta_{3+\ep}\Bigl(\frac{v}{\tau},-\frac{1}{\tau}\Bigr),
\end{align}
the exact partition function in \eqref{eq:ZN-det} becomes
\begin{align}
 Z_N=\left(\frac{2\pi}{g_sp}\right)^{\frac{N}{2}}e^{-\frac{N\th^2}{2g_sp}}
\det\left[e^{\frac{g_s}{2p}(i+j-N-1)^2}\vartheta_{3+\ep}\Bigl(\frac{i+j-N-1}{p}-\frac{\ri\th}{g_sp},\frac{2\pi\ri}{g_sp}\Bigr)\right].
\label{eq:ZN-dual}
\end{align}
Then, plugging the series expansion of Jacobi theta function
\begin{align}
 \vartheta_{3+\ep}\Bigl(\frac{i+j-N-1}{p}-\frac{\ri\th}{g_sp},\frac{2\pi\ri}{g_sp}\Bigr)=\sum_{m\in\mathbb{Z}}(-1)^{\ep m}e^{-\frac{2\pi^2m^2}{g_sp}+\frac{2\pi m\th}{g_sp}}e^{\frac{2\pi\ri m}{p}(i+j-N-1)}
\label{eq:Jacobi-m}
\end{align}
into the determinant of \eqref{eq:ZN-dual}, 
$Z_N$ is written as a sum over integer vectors $\vec{m}=(m_1,\cdots,m_N)$ 
\begin{align}
 Z_N=\sum_{\vec{m}\in\mathbb{Z}^N}\Om({\vec{m}})\exp\left(-\frac{2\pi^2}{g_sp}\sum_{i=1}^Nm_i^2
+\frac{2\pi \th}{g_sp}\sum_{i=1}^Nm_i\right),
\label{eq:Zw}
\end{align}
with some coefficient $\Om({\vec{m}})$.
This $S$-dual expression has a natural interpretation as the instanton expansion.
As in the case of undeformed Yang-Mills theory,
the instanton in question is a classical solution of gauge field
where the Dirac monopole configuration
is embedded in the Cartan part of gauge field \cite{Witten:1992xu,Witten:1991we}.  
In the weak coupling phase,
the most dominant contribution comes from the zero-instanton sector
since the instanton contribution 
$\vec{m}\ne0$ in \eqref{eq:Zw} is suppressed by the factor
$\mathcal{O}(e^{-1/g_s})$ which is non-perturbative
in $g_s$.
By setting $\vartheta_{3+\ep}=1$ in \eqref{eq:ZN-dual}
(or taking the $m=0$ term in \eqref{eq:Jacobi-m}),
the perturbative part of $Z_N$ in the weak coupling phase
is found to be
\begin{align}
\begin{aligned}
 Z_{\text{weak}}&=\left(\frac{2\pi}{g_sp}\right)^{\frac{N}{2}}e^{-\frac{N\th^2}{2g_sp}}
\det\left[e^{\frac{g_s}{2p}(i+j-N-1)^2}\right]\\
&=\left(\frac{2\pi}{g_sp}\right)^{\frac{N}{2}}e^{-\frac{N\th^2}{2g_sp}+\frac{g_s(N^3-N)}{12p}} 
\prod_{k=1}^{N-1}
\left(2\sinh\frac{g_sk}{2p}\right)^{N-k}.
\end{aligned}
\label{eq:Zweak}
\end{align}
This is exactly the same as the partition function of pure Chern-Simons theory
on $S^3$ up to a rescaling of the coupling $g_s\to g_s/p$.
The genus expansion of \eqref{eq:Zweak} is easily found from the known 
result of pure Chern-Simons theory (see e.g. \cite{Marino:2004eq})
\begin{align}
\begin{aligned}
 F_0(t)&=p^2\left[\frac{t^3}{6p^3}-\frac{\pi^2 t}{6p}-\text{Li}_3(e^{-\frac{t}{p}})+\zeta(3)\right],\\
F_1(t)&=-\frac{t}{8p}-\frac{1}{12}\log(1-e^{-\frac{t}{p}})
+\frac{t}{g_s}\log\frac{2\pi}{g_s}+\zeta'(-1)+\frac{1}{12}\log \frac{g_s}{p},\\
F_{g\geq2}(t)&=p^{2-2g}\frac{B_{2g}}{2g(2g-2)!}\left[\text{Li}_{3-2g}(e^{-\frac{t}{p}})
+\frac{B_{2g-2}}{2g-2}\right].
\end{aligned}
\end{align}

\subsection{Phase transition at $\th=0$}
As shown in \cite{Jafferis:2005jd,Arsiwalla:2005jb,Caporaso:2005ta},
the $q$YM on $S^2$ at $\th=0$ has a third order phase transition. 
The mechanism of the phase transition is essentially the same as the undeformed case
found by Douglas and Kazakov \cite{Douglas:1993iia} where
the phase transition occurs when the eigenvalue density saturates
the bound $\rho(h)\leq1$.

As we have seen in the previous subsection, the zero-instanton sector of the
weak coupling phase is
described by the pure Chern-Simons theory.
From the known eigenvalue density of Chern-Simons matrix model
\cite{Marino:2004eq}
\begin{align}
 \rho(h)=\frac{p}{\pi}\arccos\Bigl(e^{-\frac{t}{2p}}\cosh\frac{th}{2}\Bigr)
\label{eq:rho}
\end{align}
the critical point $t=t_c$ is determined by the condition $\rho(0)=1$, i.e.
\begin{align}
\frac{p}{\pi}\arccos\Bigl(e^{-\frac{t_c}{2p}}\Bigr)=1,
\end{align}
which leads to the result \eqref{eq:tc} found in \cite{Jafferis:2005jd,Arsiwalla:2005jb,Caporaso:2005ta}.

\begin{figure}[htb]
\centering
\includegraphics[width=12cm]{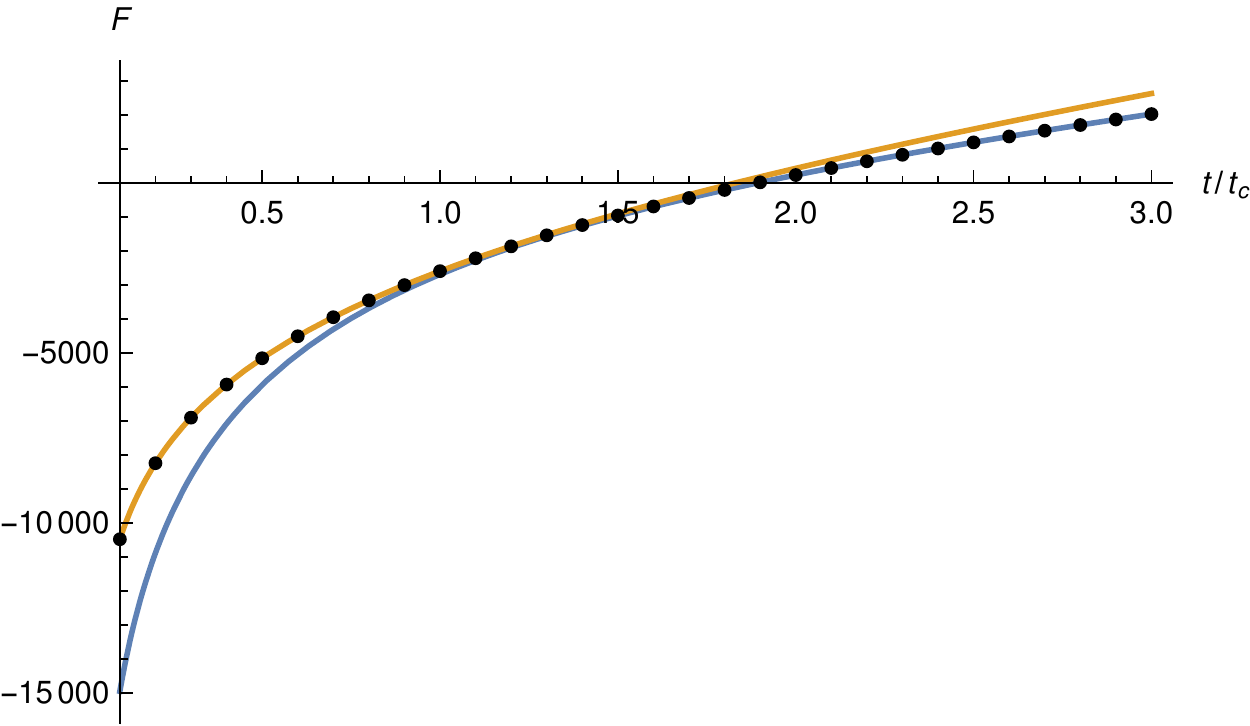}
  \caption{
Plot of free energy $F=\log Z_N$ for $p=3, N=80$ at $\th=0$.
The  dots are the numerical values of the exact partition function
\eqref{eq:ZN-det}, while the orange curve and the blue curve
represent  $\log Z_{\text{weak}}$ in \eqref{eq:Zweak} and $\log Z_{\text{gnd}}$ in \eqref{eq:Zgnd}, respectively.
}
  \label{fig:logZ}
\end{figure}

Now we can numerically study the behavior of free energy at $\th=0$
using the exact partition function \eqref{eq:ZN-det} 
at finite $N$. The determinant in \eqref{eq:ZN-det}
can be evaluated numerically with high precision
and we can plot the free energy as a function of $t=g_sN$ by varying the coupling $g_s$
with fixed $N$.
In Fig.~\ref{fig:logZ} we show the plot of free energy for $p=3,N=80$ at $\th=0$.
As we can see from this figure, the behavior of the free energy changes at $t=t_c$
from  $Z_{\text{weak}}$ \eqref{eq:Zweak} in the weak coupling phase to 
$Z_{\text{gnd}}$ \eqref{eq:Zgnd} in the strong coupling phase,
as expected from the result in 
\cite{Jafferis:2005jd,Arsiwalla:2005jb,Caporaso:2005ta}.

\section{Instanton in the weak coupling phase}\label{sec:inst}
It is argued in \cite{Jafferis:2005jd,Arsiwalla:2005jb,Caporaso:2005ta}
that the phase transition 
of $q$YM is induced by instantons.
As we have seen in the previous section  \ref{sec:weak},
the instanton expansion in the weak coupling phase
naturally arises after performing the modular $S$-transformation
of Jacobi theta function \eqref{eq:Zw}.

It turns out that when $0<\th<\pi$ the dominant contribution comes from
the $m=0$ and $m=1$ terms in the expansion of Jacobi theta function \eqref{eq:Jacobi-m}.
If we keep 
only those terms
in the expression of $Z_N$ in \eqref{eq:ZN-dual}, 
the partition function is approximated as
\begin{align}
 Z_N\approx\mathcal{N}\det(A+\xi B),
\label{eq:Z-approx}
\end{align}
where $\xi$ is the weight factor of 1-instanton 
\begin{align}
 \xi=e^{\frac{2\pi\th-2\pi^2}{g_sp}},
\end{align}
$\mathcal{N}$ is the overall factor
\begin{align}
 \mathcal{N}=\left(\frac{2\pi}{g_sp}\right)^{\frac{N}{2}}e^{-\frac{N\th^2}{2g_sp}},
\end{align}
and $A$ and $B$ in \eqref{eq:Z-approx} are the following
$N\times N$ matrices:
\begin{align}
 \begin{aligned}
  A_{i,j}&=q^{\frac{1}{2}(i+j-N-1)^2},\\
B_{i,j}&=(-1)^{N-1}x^{i+j-N-1}A_{i,j},\qquad (i,j=1,\cdots,N).
 \end{aligned}
\label{eq:AB}
\end{align}
Here we have introduced the notation $q$ and $x$ by 
\begin{align}
 q=e^{-\frac{g_s}{p}},\quad
x=e^{\frac{2\pi\ri}{p}}.
\end{align}
The validity of this approximation
\eqref{eq:Z-approx}
will be discussed in detail in the next section \ref{sec:phase}.
Note that the sign $(-1)^{N-1}=(-1)^\ep$ of $B_{i,j}$ in \eqref{eq:AB} 
comes from the sign $(-1)^{\ep m}$
in the expansion of Jacobi theta function  \eqref{eq:Jacobi-m} with $m=1$.
In this notation, $Z_{\text{weak}}$ in \eqref{eq:Zweak}
is written as
\begin{align}
 Z_{\text{weak}}=\mathcal{N}\det A.
\end{align}

Now we can define the instanton part of
partition function
by dividing $Z_N$ by 
$Z_{\text{weak}}$.
In the approximation \eqref{eq:Z-approx}
the instanton 
partition function becomes
\begin{align}
\begin{aligned}
Z_{\text{inst}}=\frac{\det(A+\xi B)}{\det A}=\det(1+\xi M),
\end{aligned}
\label{eq:Zinst}
\end{align}
where the matrix $M$ is given by
\begin{align}
 M=A^{-1}B.
\end{align}
From the explicit form of $A$ and $B$ in \eqref{eq:AB},
we find that the matrix element of $M$ 
has a simple expression
\begin{align}
 M_{i,j}=(-1)^{N-1+j-i}\frac{(x^{-1}q;q)_{i-1}}{(q;q)_{i-1}}\frac{(xq;q)_{N-j}}{(q;q)_{N-j}}
\frac{x-1}{x-q^{j-i}}x^{2j-N-1}q^{\hf(j-i)(N+2-i-j)},
\label{eq:Mmat-q}
\end{align}
where $(a;q)_k$ denotes the $q$-Pochhammer symbol
\begin{align}
 (a;q)_k=\prod_{n=0}^{k-1}(1-aq^n).
\end{align}
We have checked this 
relation \eqref{eq:Mmat-q} for $N\leq 10$, and we believe that this is true for all
$N$. In what follows we will assume that \eqref{eq:Mmat-q} holds for all $N$.
It would be interesting to find a general proof of \eqref{eq:Mmat-q}.

One can expand $Z_{\text{inst}}$
in \eqref{eq:Zinst} as a power series in $\xi$
\begin{align}
 Z_{\text{inst}}=\sum_{k=0}^\infty Z_k,
\label{eq:Zk}
\end{align} 
where $Z_0=1$ and $Z_k\propto \xi^k$.
For instance, the 1-instanton
term is given by 
\begin{align}
 Z_1=\xi\Tr M.
\label{eq:Z1inst}
\end{align}
Higher instanton corrections $Z_{k\geq2}$
will be studied in section \ref{sec:phase}.
In the 't Hooft limit,
we expect that the 1-instanton correction
$Z_1$ at $\th=0$ 
is characterized by the instanton action $S_{\text{inst}}(t)$
computed in  \cite{Jafferis:2005jd,Arsiwalla:2005jb,Caporaso:2005ta}
\begin{align}
\xi_0 \Tr M\sim 
e^{-\frac{1}{g_s}S_{\text{inst}}(t)},
\label{eq:inst-scale}
\end{align}
with $\xi_0$ being 
\begin{align}
 \xi_0=\xi_{\th=0}=e^{-\frac{2\pi^2}{g_sp}}.
\end{align}
In \cite{Jafferis:2005jd}, it was found that
the instanton action 
is given by the integral of 
eigenvalue density $\rho(h)$ in \eqref{eq:rho}
along the {\it imaginary axis}
\begin{align}
 S_{\text{inst}}(t)=2\pi t\int_{0}^{h_0}dh\Bigl[1-\rho(\ri h)\Bigr].
\label{eq:Sinst}
\end{align}
Here the upper bound of integral, $h_0$, is determined by the condition $\rho(\ri h_0)=1$:
\begin{align}
 \rho(\ri h)=\frac{p}{\pi}\arccos\Bigl(e^{-\frac{t}{2p}}\cos\frac{th}{2}\Bigr),\quad
h_0=\frac{2}{t}\arccos\Bigl(e^{\frac{t}{2p}}\cos\frac{\pi}{p}\Bigr).
\end{align}
This suggests that the large $N$
limit of instanton in the weak coupling phase can be 
thought of as a complex instanton.
A similar phenomenon was observed in the GWW model as well
\cite{Marino:2008ya,Buividovich:2015oju}.

\begin{figure}[htb]
\centering
\includegraphics[width=12cm]{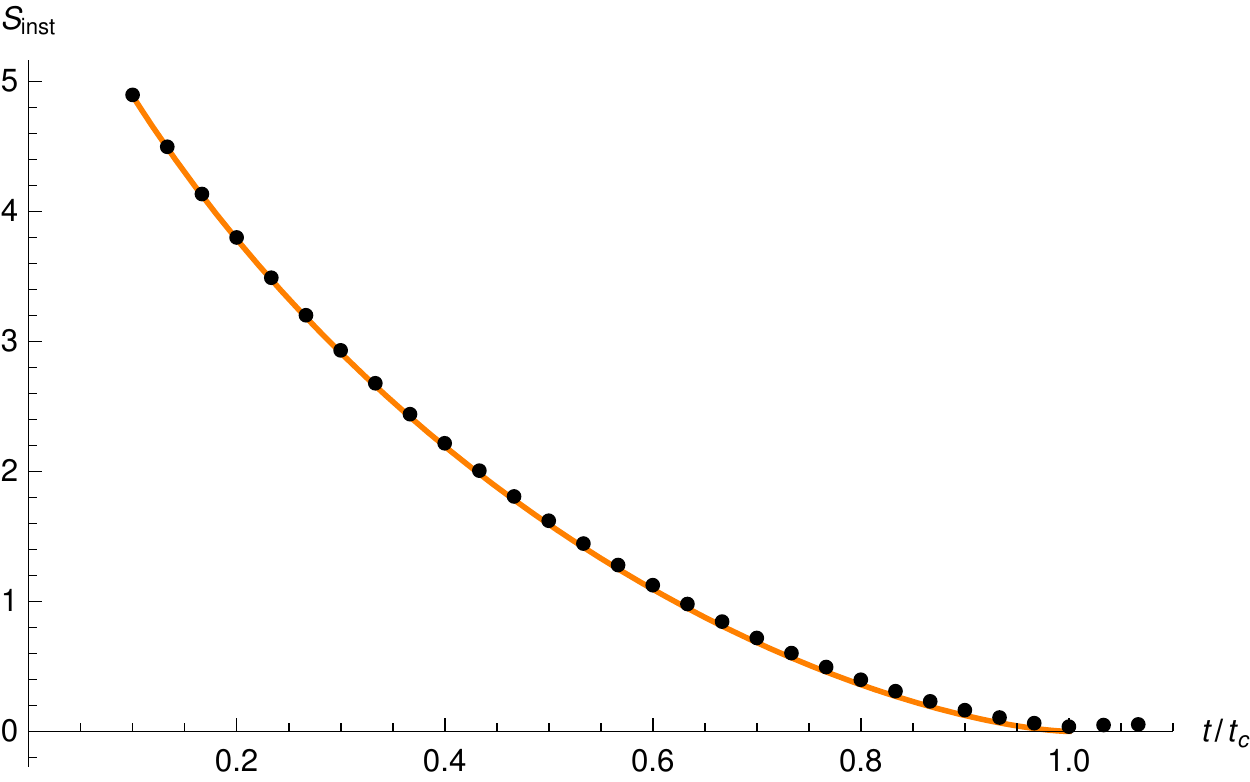}
  \caption{
Plot of instanton action at $\th=0$.
The dots are the numerical value of $-g_s\log(\xi_0\Tr M)$ for $p=3, N=400$,  
while the orange curve 
represents $S_{\text{inst}}(t)$ in \eqref{eq:Sinst}.
}
  \label{fig:sinst}
\end{figure}

In Fig.~\ref{fig:sinst} we show the plot of $\xi_0\Tr M$
for $p=3,N=400$ using the exact form of $M$ in \eqref{eq:Mmat-q}.
One can clearly see that the exact result of $M$
nicely reproduces the analytic form of 
instanton action in \eqref{eq:Sinst}.
The instanton action vanishes at $t=t_c$
as shown in \cite{Jafferis:2005jd,Arsiwalla:2005jb,Caporaso:2005ta},
which is also reproduced numerically by our exact result of $M$.
This leads to a physical picture of the phase transition
that it is triggered by the condensation of instantons, as in the case of undeformed theory
\cite{Gross:1994mr}.  
We also observed numerically that the 1-instanton correction 
$\xi_0\Tr M$ is always positive for both even $N$ and odd $N$
in the weak coupling phase;
we emphasize that the sign $(-1)^{N-1}$ of $B$ in \eqref{eq:AB}
is crucial for this positivity of 1-instanton correction.

\subsection{$Z_{\text{inst}}$ in the $p\to\infty$ limit}
We expect that $Z_{\text{inst}}$
reduces to the known instanton correction of undeformed theory
in the limit $p\to\infty$ with $g=g_sp$ fixed \eqref{eq:undeform-coupling}.
Indeed, we find that the 1-instanton term \eqref{eq:Z1inst} reduces to
\begin{align}
 \lim_{p\to\infty}\Tr M=(-1)^{N-1}L_{N-1}^{1}\Bigl(\frac{4\pi^2}{g}\Bigr),
\label{eq:un-1inst}
\end{align}
where $L_{n}^{\al}(x)=\frac{1}{n!}x^{-\al}e^x \del_x^n(x^{n+\al}e^{-x})$ denotes
the Laguerre polynomial.
As expected, this agrees with the result of 1-instanton correction of undeformed theory
\cite{Gross:1994mr}.
This implies that the 1-instanton term $\Tr M$ in $q$YM
can be thought of as a certain $q$-deformation of the Laguerre polynomial.
As shown in \cite{Gross:1994mr}, the large $N$  limit of
$L_{N-1}^{1}(4\pi^2/g)$ has a sign $(-1)^{N-1}$ which
is precisely canceled by the overall sign in \eqref{eq:un-1inst}
coming from the modular $S$-transformation of
Jacobi theta function.
The resulting 1-instanton factor \eqref{eq:un-1inst} in the undeformed theory
is always positive for both even $N$ and odd $N$, which is consistent with the above
observation of the positivity of $\Tr M$ in the $q$-deformed theory. 

More generally, we find
\begin{align}
 \lim_{p\to\infty}\det(1+\xi M)=
\det(1+\xi \til{M}),
\label{eq:laguerre-det}
\end{align}
where the $N\times N$ matrix $\til{M}$ is given by
\begin{align}
 \til{M}_{i,j}=(-1)^{N-1}L_{i-1}^{j-i}\Bigl(\frac{4\pi^2}{g}\Bigr),\quad (i,j=1,\cdots,N).
\label{eq:tilM}
\end{align}
The right hand side of \eqref{eq:laguerre-det}
is exactly the generating function of the expectation value of 't Hooft loops
in the anti-symmetric representations in 4d $\mathcal{N}=4$ $U(N)$
super Yang-Mills theory\footnote{
This generating function of 't Hooft loops 
is obtained from the
exact result of Wilson loops in \cite{Fiol:2013hna}
by the S-duality of 4d $\mathcal{N}=4$ super Yang-Mills
theory.}, up to a change of sign of the coupling $g=-g_{4d}^2$ 
\cite{Giombi:2009ek}.
This is expected since the computation of 't Hooft loops
in 4d $\mathcal{N}=4$ super Yang-Mills theory localizes to  2d Yang-Mills theory on $S^2$ in 
the instanton sector \cite{Giombi:2009ek}.

We have checked that \eqref{eq:un-1inst} and \eqref{eq:laguerre-det}
hold for small $N$, but we do not have a general proof.
It would be nice to find a proof of \eqref{eq:un-1inst} and \eqref{eq:laguerre-det}
for general $N$.
\section{Phase diagram of $q$-deformed Yang-Mills theory at $\th\ne0$}\label{sec:phase}
The phase diagram of $q$YM at non-zero $\th$ was conjectured in
\cite{Jafferis:2005jd}.
In this section we will examine this conjecture using our exact result of 
partition function at finite $N$.

We first notice that using the symmetry of the exact partition function
\begin{align}
 \th\to -\th,\qquad \th\to\th+2\pi,
\label{eq:sym}
\end{align}
we can restrict $\th$ to the region $0\leq\th\leq\pi$
without loss of generality.
As discussed
in \cite{Witten:1998uka}, for general value of $\th$
we should minimize the energy of $Z_{\text{weak}}$ in \eqref{eq:Zweak}
under all $2\pi$-shifts of $\th$
\begin{align}
 \underset{\ell \in\mathbb{Z}}{\text{min}}~\frac{N(\th+2\pi \ell)^2}{2g_sp}.
\end{align}
The minimum is given by $\ell=0$ 
if $\th$ is in the range $0\leq\th\leq\pi$,
and hence we can safely use the $S$-dual expression of $Z_N$ in
\eqref{eq:ZN-dual} and \eqref{eq:Z-approx} in this region of $\th$.
Thus the 1-instanton term for $0\leq\th\leq\pi$ is given by
\begin{align}
 Z_1=\xi\Tr M=e^{\frac{2\pi\th}{g_sp}}\xi_0\Tr M\sim e^{\frac{2\pi\th}{g_sp}-\frac{1}{g_s}S_{\text{inst}}(t)},
\label{eq:Z1-thne0}
\end{align}
where we used \eqref{eq:inst-scale}.
Then the critical value $t=t_*(\th)$ where $Z_1$ becomes of order one
 is determined by the condition that the exponent in \eqref{eq:Z1-thne0}
vanishes
\begin{align}
 S_{\text{inst}}\bigl(t_*(\th)\bigr)=\frac{2\pi\th}{p}.
\label{eq:t*}
\end{align}
It is conjectured in \cite{Jafferis:2005jd} that 
the critical line $t=t_*(\th)$ on the $t\text{-}\th$ plane 
is just the first one of such critical lines;
there are many critical lines on the $t\text{-}\th$ plane
which accumulate at $(t,\th)=(0,\pi)$ (see Figure 3 in \cite{Jafferis:2005jd}).

\begin{figure}[htb]
\centering
\subcaptionbox{Plot of $r$\label{sfig:r}}{\includegraphics[width=7cm]{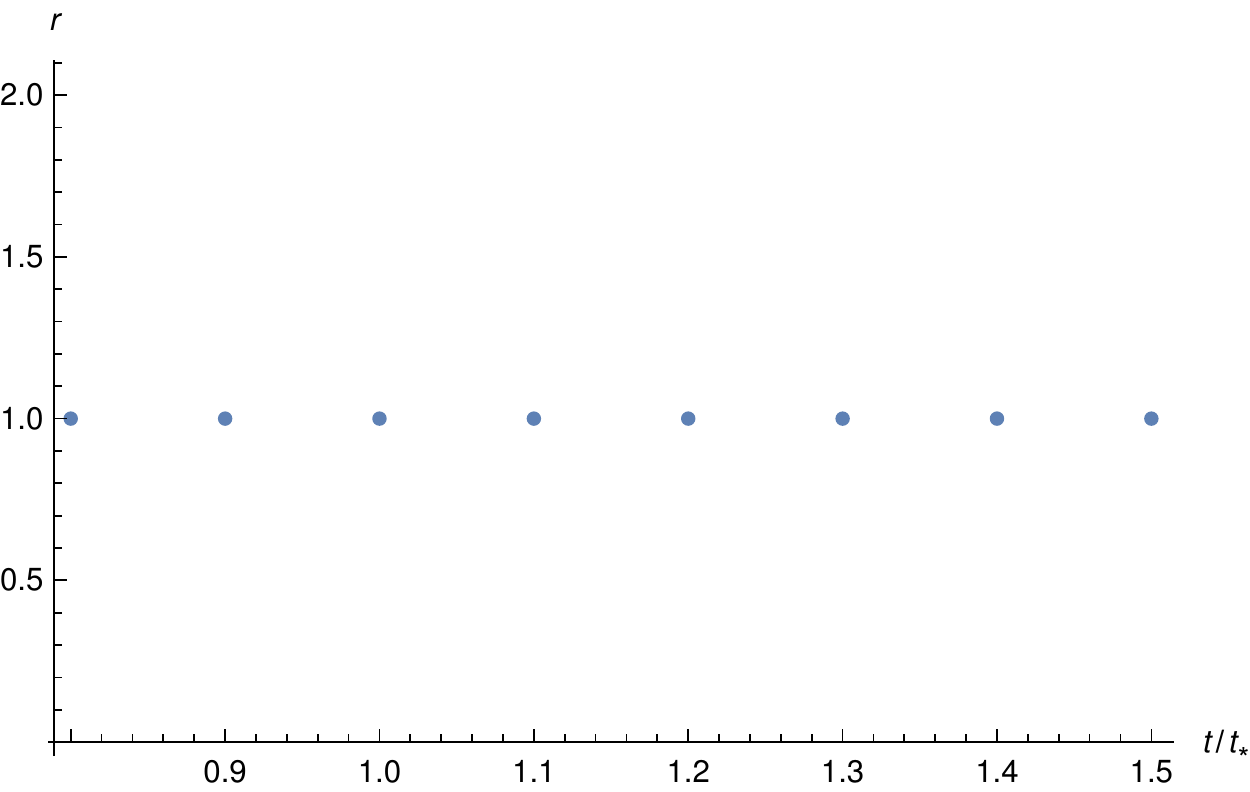}}
\hskip9mm
\subcaptionbox{Plot of $|r-1|$\label{sfig:r-1}}{\includegraphics[width=7cm]{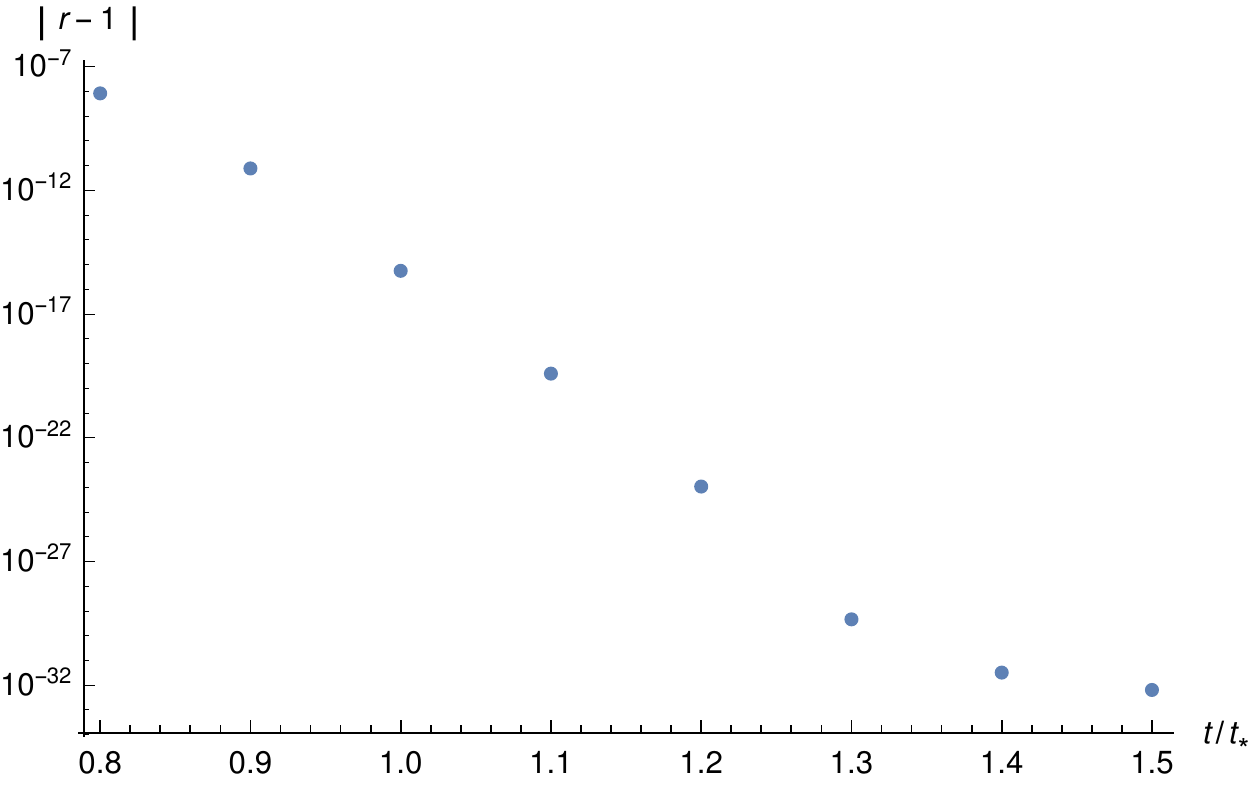}}
  \caption{
Plot of the ratio $r$ \eqref{eq:Z-ratio}  for $p=3,N=80$ at $\th=\frac{\pi}{3}$.
The horizontal axis is $t/t_*(\pi/3)$ where $t_*(\th)$ is defined by \eqref{eq:t*}.}
  \label{fig:ratio}
\end{figure}

The conjectured phase diagram of \cite{Jafferis:2005jd} is based on the following 
two assumptions:
\begin{itemize}
 \item[(i)] Only the instantons with charges $\vec{m}=(1,1,\cdots,1,0\cdots,0)$ 
in the expansion \eqref{eq:Zw} are relevant for the phase transition at $\th\ne0$.
\item[(ii)] There is a series of phase transitions at $\th\ne0$
where the instantons of the type in (i) exchange dominance. 
\end{itemize}
The assumption (i) amounts to using the approximation in 
\eqref{eq:Z-approx}.
We can test this assumption by computing the following ratio
numerically 
\begin{align}
 r=\frac{Z_{\text{weak}}Z_{\text{inst}}}{Z_N},
\label{eq:Z-ratio}
\end{align}
where $Z_{\text{weak}}$ and $Z_{\text{inst}}$ are given by
\eqref{eq:Zweak} and \eqref{eq:Zinst}, respectively.
In Fig.~\ref{fig:ratio}, we show the plot of this ratio for $p=3,N=80$ at $\th=\frac{\pi}{3}$.
From this figure, one can see that this ratio is very close to 1
\begin{align}
 r\approx 1,
\label{eq:r1}
\end{align}
which confirms the assumption (i).

\begin{figure}[htb]
\centering
\subcaptionbox{$F_{\text{inst}}$\label{sfig:Finst}}{\includegraphics[width=7cm]{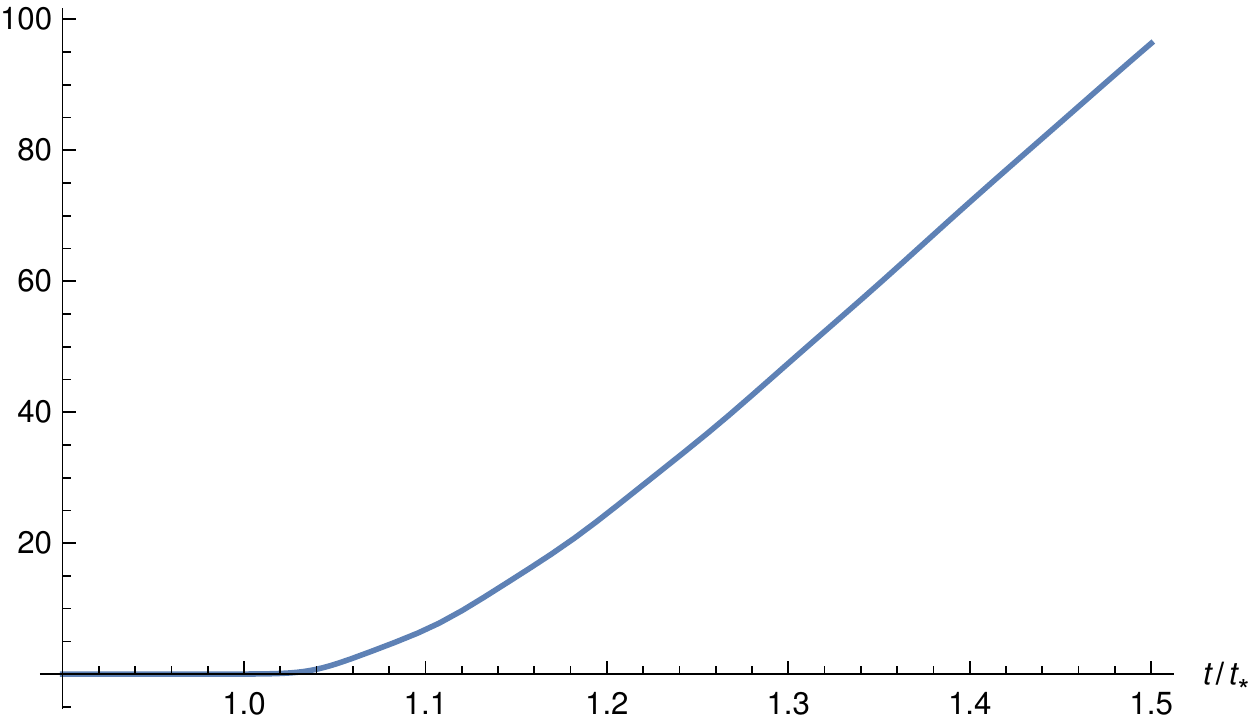}}
\hskip9mm
\subcaptionbox{$\del_t F_{\text{inst}}$\label{sfig:d1F}}{\includegraphics[width=7cm]{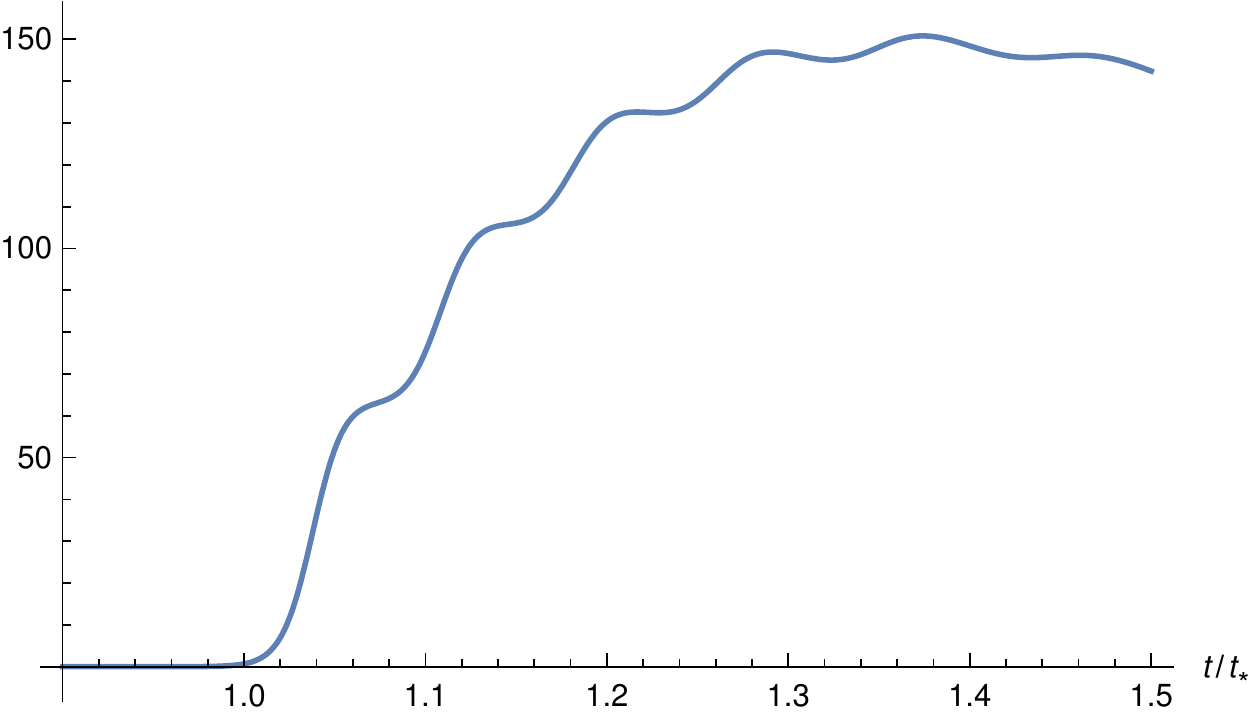}}\\
\subcaptionbox{$\del_t^2F_{\text{inst}}$\label{sfig:d2F}}{\includegraphics[width=7cm]{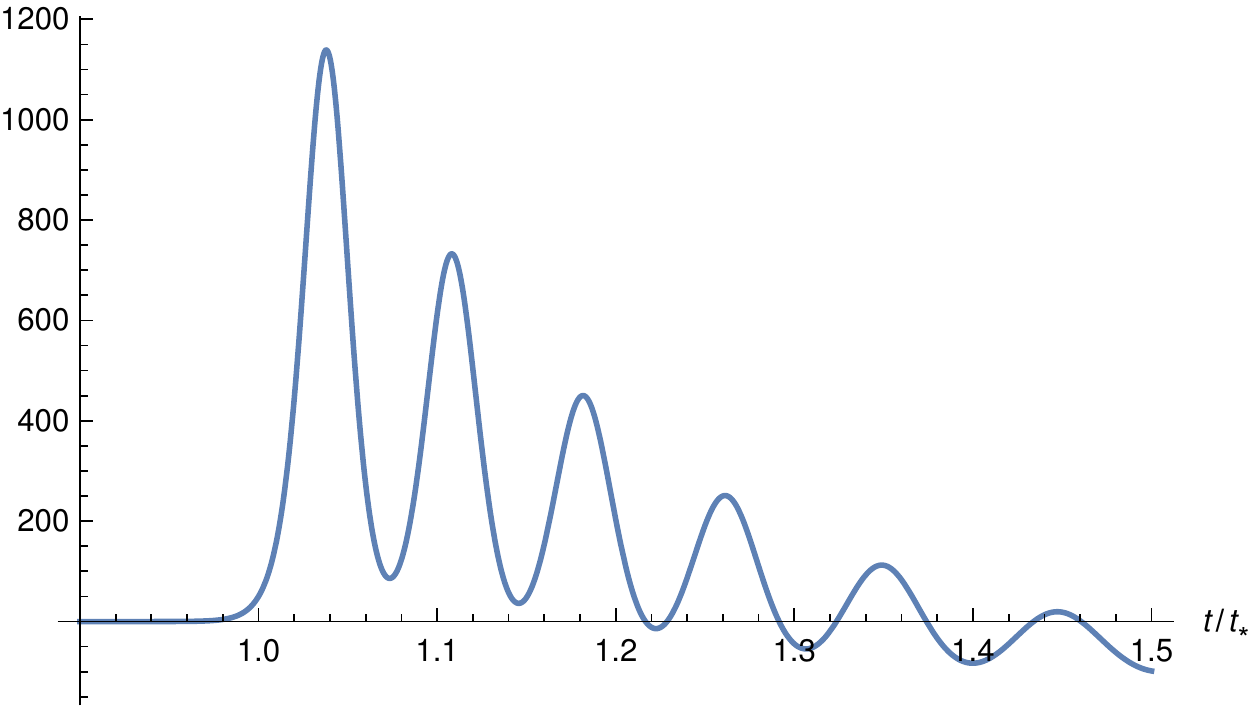}}
\hskip9mm
\subcaptionbox{$\del_t^3 F_{\text{inst}}$\label{sfig:d3F}}{\includegraphics[width=7cm]{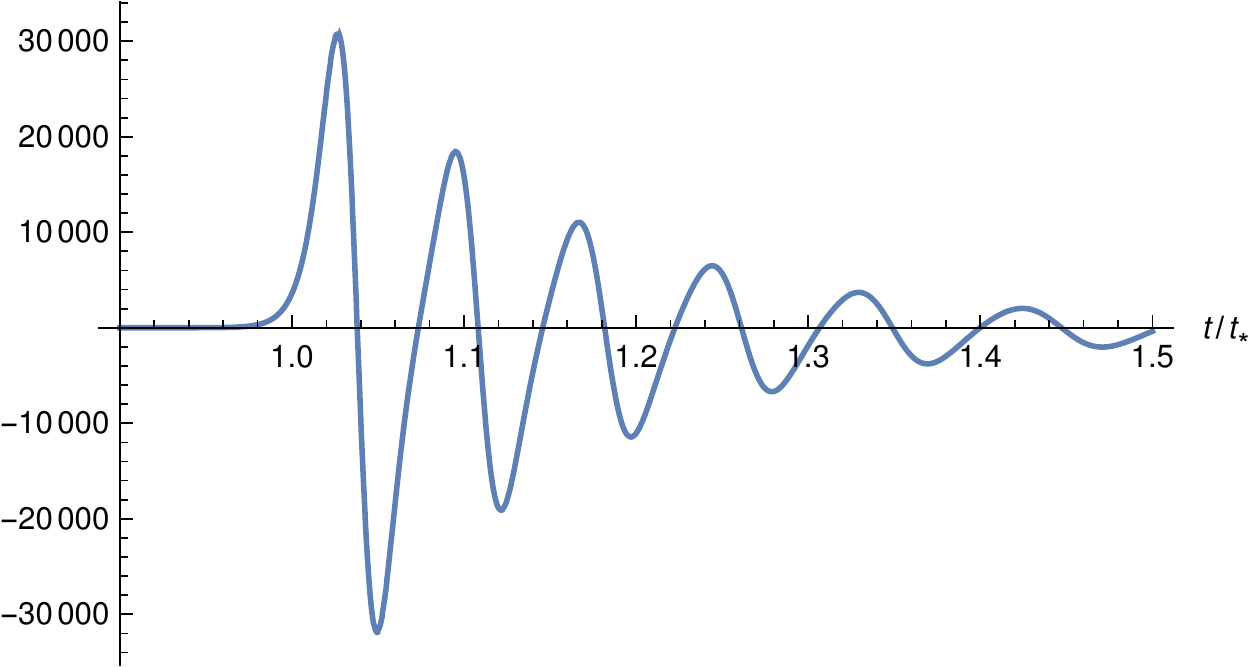}}
  \caption{
Plot of  $F_{\text{inst}}=\log Z_{\text{inst}}$ and its derivatives
for $p=3, N=80$ at $\th=\frac{\pi}{3}$
}
  \label{fig:Finst}
\end{figure}

Next consider the assumption (ii). In Fig.~\ref{fig:Finst}, we show the plot of
the instanton part of free energy $F_{\text{inst}}=\log Z_{\text{inst}}$ and its derivatives
for $p=3, N=80$ at $\th=\frac{\pi}{3}$.
\footnote{To draw this plot, we first compute 
$F_{\text{inst}}(t)$ numerically at discrete values of $t$, and then find the interpolating function
from the discrete data. In Fig.~\ref{sfig:d1F}-\ref{sfig:d3F}, we plot the derivatives of this interpolating function.}
One can see that the third derivative $\del_t^3F_{\text{inst}}$ 
in Fig.~\ref{sfig:d3F} has
several jumps at different values of $t$, and the first jump (or discontinuity of 
$\del_t^3F_{\text{inst}}$) occurs 
at $t\approx t_*(\th)$.
This is consistent with the assumption (ii).
In Fig.~\ref{fig:Finst} we have used the approximate instanton partition function $Z_{\text{inst}}$
in \eqref{eq:Zinst}, but the plot of the exact partition function
$Z_N/Z_{\text{weak}}$ does not change much from Fig.~\ref{fig:Finst} due to
the property $r\approx1$ \eqref{eq:r1}.
In Fig.~\ref{sfig:d3F}, the third derivative 
$\del_t^3F_{\text{inst}}$ has sharp discontinuities only at 
the 
first few zeros of $\del_t^3F_{\text{inst}}$, but we expect that this is a finite $N$
effect and in the strict large $N$ limit they become sharp phase transitions.

\begin{figure}[htb]
\centering
\subcaptionbox{Plot of $\displaystyle \frac{\sum_{n=0}^k Z_n}{Z_{\text{inst}}},~~(k=0,\cdots,4)$\label{sfig:zinst}}{\includegraphics[width=7cm]{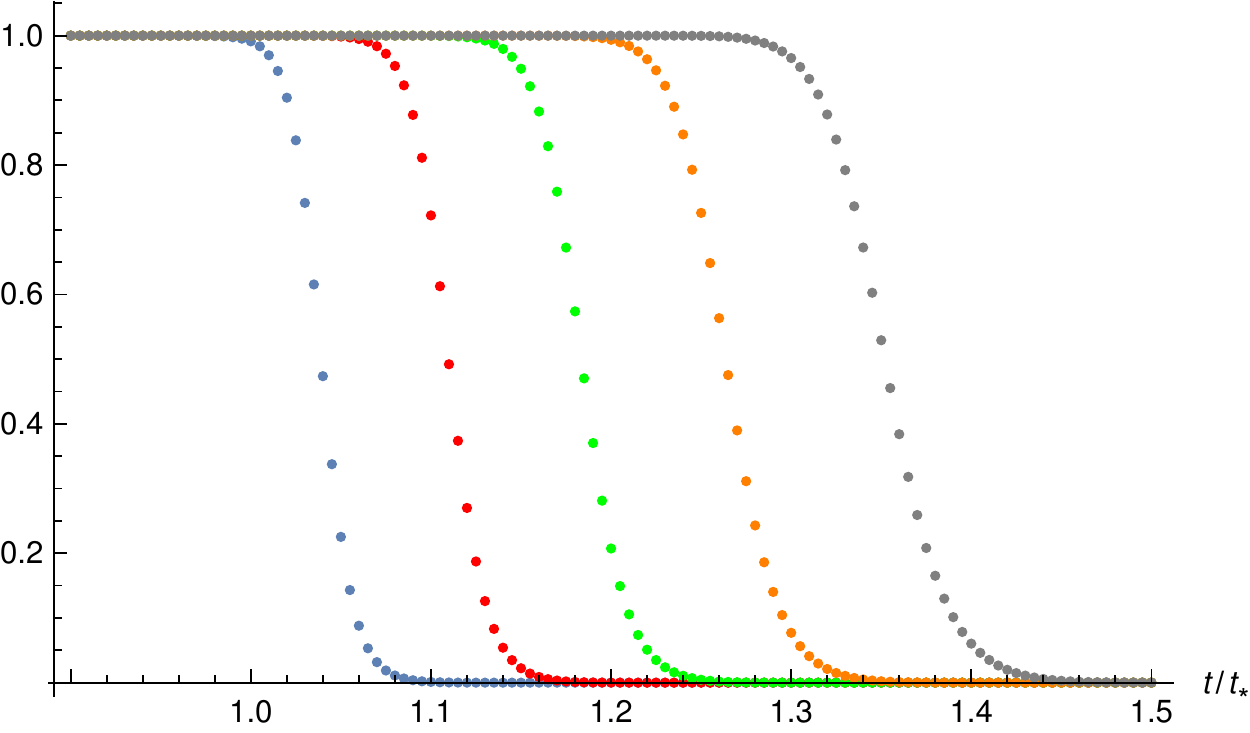}}
\hskip9mm
\subcaptionbox{Plot of $\dfrac{Z_k}{Z_{\text{inst}}},~~(k=1,\cdots,4)$\label{sfig:zinst-zk}}{\includegraphics[width=7cm]{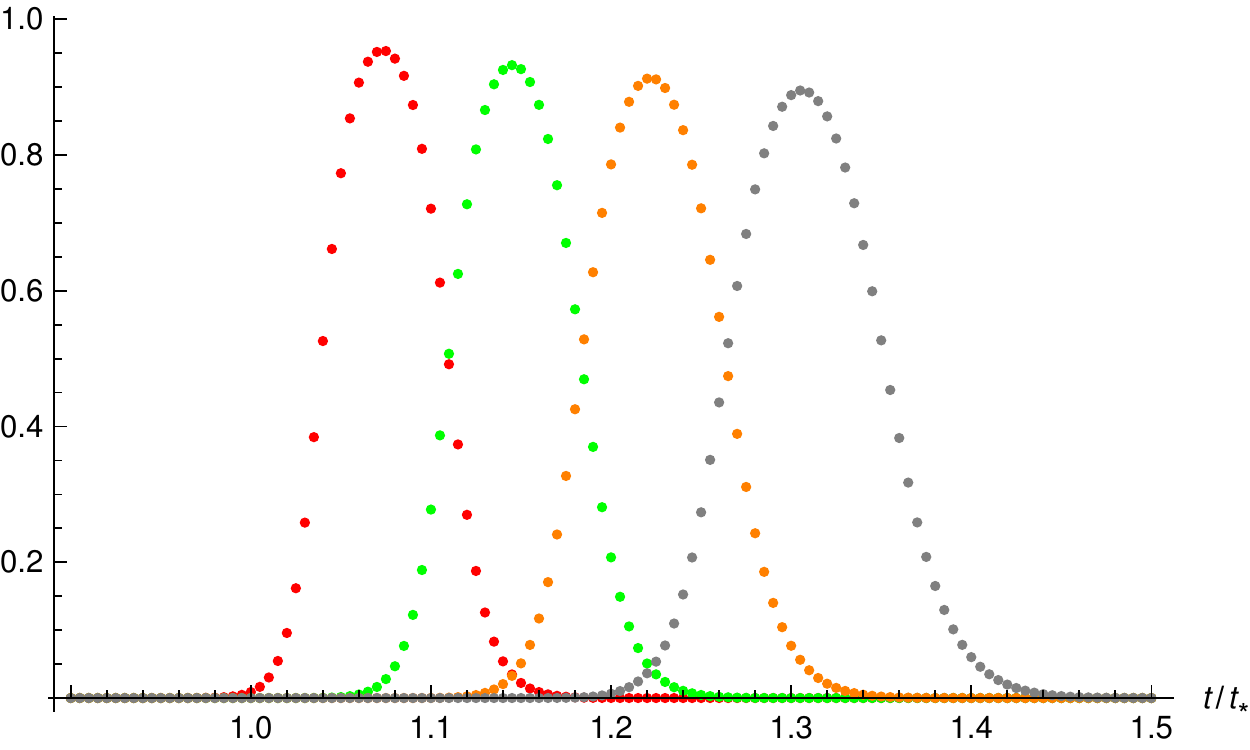}}
  \caption{
Plot of  $k$-instanton contribution for $p=3, N=80$ at $\th=\frac{\pi}{3}$.
In \subref{sfig:zinst} 
we show the sum of instanton contributions up to $k$-instantons normalized by 
$Z_{\text{inst}}$, and in \subref{sfig:zinst-zk} we show each $k$-instanton contribution 
individually. In both figures  \subref{sfig:zinst} and \subref{sfig:zinst-zk},
we used the same colors: 
$k=1$ ({\color{red}red}), $k=2$ ({\color{green}green}), $k=3$ ({\color{orange}orange}), 
$k=4$ ({\color{gray}gray}).
}
  \label{fig:Zinst-ratio}
\end{figure}

We can collect more evidence for the 
assumption (ii) by 
evaluating the $k$-instanton contribution $Z_k$ in \eqref{eq:Zk} separately.
In Fig.~\ref{fig:Zinst-ratio}, we show the plot of 
the $k$-instanton contribution ($k=1,\cdots,4$) for $p=3,N=80$ at $\th=\frac{\pi}{3}$.
One can see that the dominant instanton changes from $k=1$ to $k=4$ as $t$ increases, 
and the exchange of dominance occurs at different values of $t$ for different instanton number 
$k$. In other words, there is a series of phase transitions at $t=t_k~(k=1,2,\cdots)$
where $Z_{k-1}$ and $Z_k$ exchange dominance at $t=t_k$. 
Our numerical result in Fig.~\ref{fig:Zinst-ratio} gives strong evidence for the assumption (ii).
Also, we observed numerically that
the difference of $t_k$ and $t_{k+1}$ decreases as $N$ becomes large,
and the difference scales approximately as $1/N$
\begin{align}
 t_{k+1}-t_k\sim 1/N.
\label{eq:tk-scale}
\end{align}
This suggests that the $1/N$ correction of instanton factor,
in particular the {\it prefactor} $f_k$ of $k$-instanton,
is important for the understanding of 
the scaling behavior \eqref{eq:tk-scale}
\begin{align}
 Z_k=f_k(t,g_s)
e^{\frac{2\pi\th k}{g_sp}-\frac{k}{g_s}S_{\text{inst}}(t)}.
\end{align}
However, we were unable to find the analytic form
of the prefactor $f_k$ and hence we could not determine
the analytic form of the critical value $t=t_k$.
In the next section, we will consider the
phase diagram of undeformed Yang-Mills theory,
where the instanton prefactor is more tractable analytically.

\begin{figure}[htb]
\centering
\includegraphics[width=12cm]{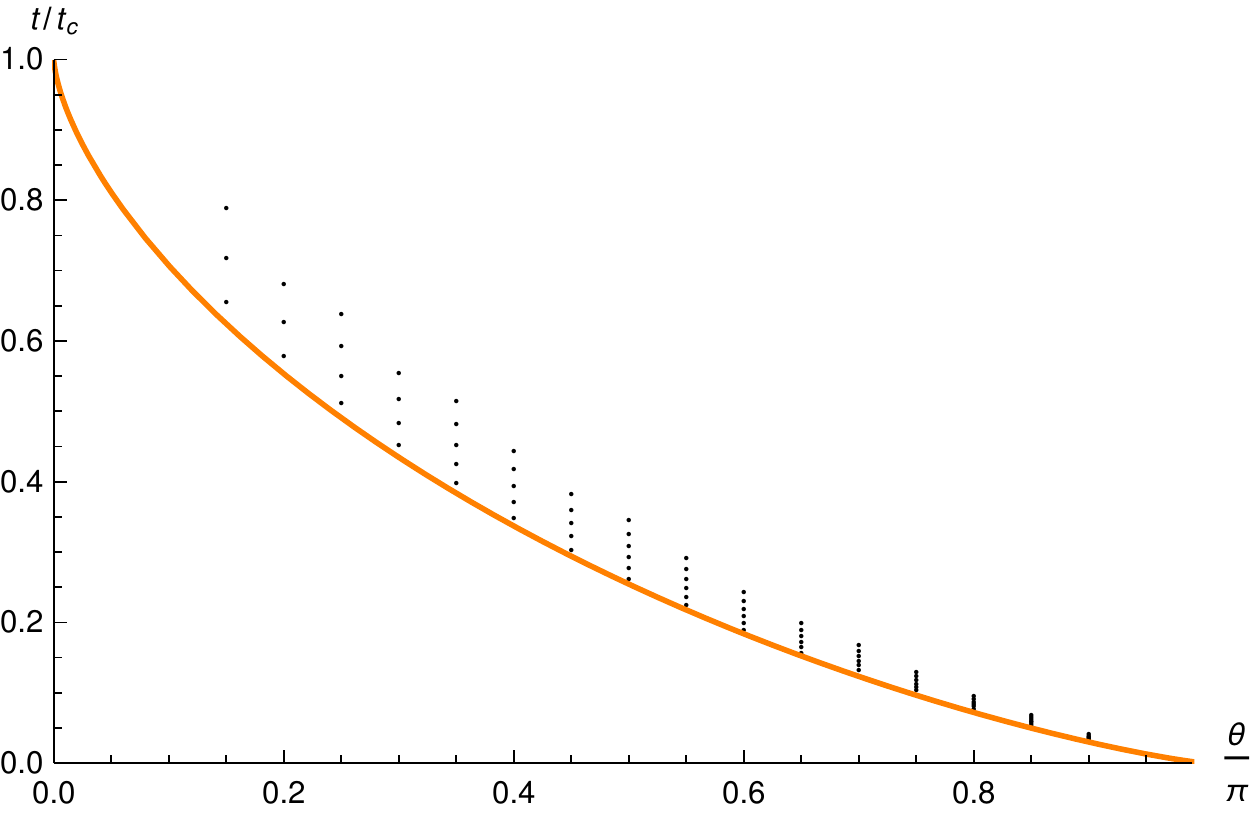}
  \caption{
Phase diagram at non-zero $\th$ ($p=3,N=80$). We plotted the local maxima of $\del_t^2 F_{\text{inst}}$
at fixed $\th$ in the range $t_*(\th)\leq t\leq 1.4t_*(\th)$ and
varied $\th$ with step $\lap\th/\pi=0.05$.
The orange curve is the line $t=t_*(\th)$ given by \eqref{eq:t*}.
}
  \label{fig:phase}
\end{figure}

Finally, we can draw the phase diagram of $q$YM by using our exact 
result of $Z_{\text{inst}}$ in \eqref{eq:Zinst}.
To do this, we first observe from Fig.~\ref{fig:Finst}
that the local maximum of the second derivative
$\del_t^2F_{\text{inst}}$ corresponds to
the (approximate) discontinuous point of the third derivative 
$\del_t^3F_{\text{inst}}$.
Based on this observation, in Fig.~\ref{fig:phase} 
we plot the local maxima of 
$\del_t^2F_{\text{inst}}$ for $p=3,N=80$ for several values of $\th$.
Our result in Fig.~\ref{fig:phase} agrees with the conjectured phase diagram in
\cite{Jafferis:2005jd}, at least qualitatively.
In particular we can see from Fig.~\ref{fig:phase} 
that the transition curves seem to
accumulate at the point 
$(t,\th)=(0,\pi)$.

\section{Phase diagram of undeformed Yang-Mills theory at $\th\ne0$}\label{sec:undeform}
In this section, we consider the phase transition curves in the undeformed theory, where
the instanton prefactor can be studied analytically.
In fact, the analytic form of the
prefactor of 1-instanton term has been already obtained in \cite{Gross:1994mr}.

Let us consider the instanton contributions 
in the undeformed Yang-Mills theory \eqref{eq:laguerre-det}
\begin{align}
 Z_{\text{inst}}=\det(1+\xi \til{M})
\label{eq:zin-un}
\end{align}
where $\til{M}$ is given by \eqref{eq:tilM}.
The coupling in the undeformed theory is defined in \eqref{eq:undeform-coupling}.
It is convenient to rescale the coupling $g$ and $A$ in \eqref{eq:undeform-coupling} as
\begin{align}
\hat{g}=\frac{g}{\pi^2}, \quad a=\hat{g}N=\frac{A}{\pi^2}.
\end{align}
In this normalization, when $\th=0$ the phase transition occurs at $a=1$, 
corresponding to the critical value $A=\pi^2$ found in \cite{Douglas:1993iia}.
In terms of these rescaled couplings,  $\til{M}$ and $\xi$ in \eqref{eq:zin-un} become
\begin{align}
 \til{M}_{i,j}=(-1)^{N-1}L_{i-1}^{j-i}\bigl(4/\hat{g}\bigr),\qquad
\xi=e^{\frac{2(\th-\pi)}{\pi\hat{g}}}.
\end{align}

By expanding $Z_{\text{inst}}$ in \eqref{eq:zin-un} as a power series in $\xi$ 
we can define the $k$-instanton term $Z_k$ as in the case of $q$YM in \eqref{eq:Zk}.
As we have seen in \eqref{eq:un-1inst},
the 1-instanton term can be written in a closed form 
\begin{align}
 Z_1=\xi\Tr\til{M}=\xi (-1)^{N-1}L_{N-1}^1\bigl(4/\hat{g}\bigr).
\label{eq:z1-lag}
\end{align}
The large $N$ limit of $Z_1$ has been studied in \cite{Gross:1994mr}
and the result reads
\begin{align}
 Z_1=f_1(a,\hat{g})e^{-\frac{1}{\hat{g}}S_{\text{inst}}(a)+\frac{2\th}{\pi\hat{g}}},
\end{align}
where the instanton action is given by\footnote{This is obtained by integrating 
the eigenvalue density of Gaussian matrix model along the imaginary axis, as in the case of $q$-deformed theory \eqref{eq:Sinst}
\begin{align}
 \begin{aligned}
  S_{\text{inst}}(a)=2\pi a\int_0^{h_0}dh\bigl[1-\rho_{\text{G}}(\ri h)\bigr],\quad
\rho_{\text{G}}(\ri h_0)=1,\quad \rho_{\text{G}}(h)=\frac{\pi a}{2}\rt{\frac{4}{\pi^2a}-h^2}.
 \end{aligned}
\end{align}}
\begin{align}
S_{\text{inst}}(a)=2\Bigl[\rt{1-a}-a\cosh^{-1}(1/\rt{a})\Bigr].
\end{align}
The prefactor $f_1(a,\hat{g})$ was also computed in \cite{Gross:1994mr}\footnote{It would be possible to compute the $\mathcal{O}(\hat{g})$ correction of $f_1(a,\hat{g})$ 
using the known asymptotic behavior of Laguerre polynomials \cite{DLMF}.}
\begin{align}
  f_1(a,\hat{g})=
\qu \left[\frac{\hat{g}^2a^2}{4\pi^2(1-a)}\right]^\qu \Bigl(1+\mathcal{O}(\hat{g})\Bigr).
\end{align}
As we mentioned in section \ref{sec:inst}, the sign $(-1)^{N-1}$ in \eqref{eq:z1-lag}
is precisely canceled by the same sign coming from the large $N$ limit of Laguerre polynomial, and the final
result of the prefactor $f_1(a,\hat{g})$ does not have this sign.
We expect that the $k$-instanton contribution $Z_k$ has the form
\begin{align}
 Z_k=f_k(a,\hat{g}) e^{-\frac{k}{\hat{g}}S_{\text{inst}}(a)+\frac{2k\th}{\pi\hat{g}}}.
\label{eq:Zk-form}
\end{align}
Namely, 
$\log Z_k\approx k\log Z_1$ at the leading order in $\hat{g}$ expansion.
We have checked this behavior numerically for $k=2,3,4$.
We would like to find the prefactor $f_k(a,\hat{g})$ for $k\geq2$ but 
we were unable to determine them analytically. However, one can study
the prefactor $f_k(a,\hat{g})$ numerically using the exact result at finite $N$.
For instance, from the numerical analysis of the
large $N$ behavior of the 2-instanton 
\begin{align}
 Z_2=\frac{\xi^2}{2}\Bigl[\big(\Tr\til{M}\big)^2-\Tr\big(\til{M}\big)^2\Bigr],
\end{align}
and assuming $f_2(a,\hat{g})=\al f_1(a,\hat{g})^\bt$ with some constants $\al,\bt$ at the leading order in $\hat{g}$ expansion,
we can determine the parameters $\al,\bt$ numerically. In this way we find the prefactor of 2-instanton
\begin{align}
f_2(a,\hat{g})=\frac{\pi}{128}\frac{\hat{g}^2a^2}{4\pi^2(1-a)}\Bigl(1+\mathcal{O}(\hat{g})\Bigr).
\end{align}
It would be interesting to derive this result analytically.

These prefactors are important to find the
critical lines at non-zero $\th$. 
The critical value $a=a_*(\th)$
at the leading order in $\hat{g}$ expansion is determined by the condition that
the exponent of $Z_k$ in \eqref{eq:Zk-form} vanishes
\begin{align}
 S_{\text{inst}}\bigl(a_*(\th)\bigr)=\frac{2\th}{\pi}.
\label{eq:a*}
\end{align}
This vanishing condition of the exponential factor is common for all $k$,
and this condition alone is not enough to distinguish the dominant $Z_k$.
It turns out that it is important to include the effect of prefactor
to explain the splitting of critical values observed in \eqref{eq:tk-scale}.
Let us consider the $\hat{g}$ correction for
the first two critical values $a_1(\th),a_2(\th)$ determined by
the condition
\begin{align}
 \frac{Z_1}{Z_0}(a_1(\th))=1,\quad
\frac{Z_2}{Z_1}(a_2(\th))=1.
\end{align}
Including the contribution of prefactors, 
we find the 
$\mathcal{O}(\hat{g})$ deviation of $a_1(\th)$ and $a_2(\th)$
from the leading term $a_*(\th)$ in \eqref{eq:a*}
 \begin{align}
  \begin{aligned}
   a_1(\th)&=a_*(\th)+\hat{g} \frac{\log f_1(a_*(\th),\hat{g})}{S'_{\text{inst}}(a_*(\th))}+\mathcal{O}(\hat{g}^2),\\
a_2(\th)&=a_*(\th)+\hat{g} \frac{\log f_2(a_*(\th),\hat{g})-\log f_1(a_*(\th),\hat{g})}{S'_{\text{inst}}(a_*(\th))}+\mathcal{O}(\hat{g}^2),
  \end{aligned}
\label{eq:a12}
 \end{align}
where $S'_{\text{inst}}(a)=\del_a S_{\text{inst}}(a)$ is given by
\begin{align}
 S'_{\text{inst}}(a)=-2\cosh^{-1}(1/\rt{a}).
\end{align}
This nicely explains the splitting $a_2-a_1\sim\mathcal{O}(1/N)$ observed in \eqref{eq:tk-scale}.
\begin{figure}[htb]
\centering
\includegraphics[width=12cm]{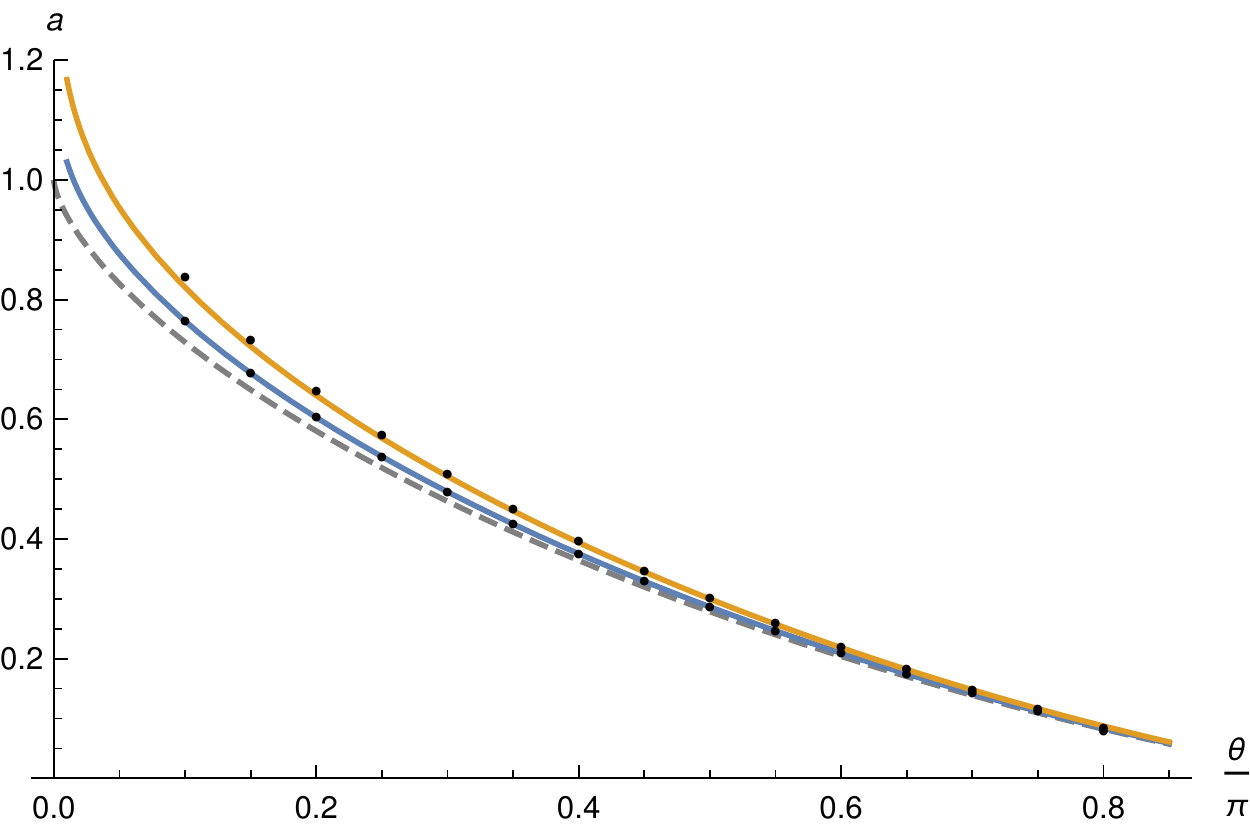}
  \caption{
The phase transition lines of undeformed Yang-Mills theory 
at non-zero $\th$ ($N=80$). The black dots are  the first two local maxima of $\del_t^2 F_{\text{inst}}$
at fixed $\th$ 
and we varied $\th$ with step $\lap\th/\pi=0.05$.
The blue  and  the orange curves are the lines $a=a_1(\th)$ and $a=a_2(\th)$, respectively.
The gray dashed curve represents the curve $a=a_*(\th)$ without taking into account
the effect of instanton prefactors.
}
  \label{fig:ac}
\end{figure}

Now we can draw the phase diagram of undeformed theory in a similar manner as 
the $q$-deformed case in Fig.~\ref{fig:phase}. 
In Fig.~\ref{fig:ac}, we plot the first two maxima of $\del_t^2 F_{\text{inst}}$
computed numerically from $Z_{\text{inst}}$ in 
\eqref{eq:zin-un}.
One can see that numerical data points fit well on the curves $a=a_1(\th)$ and $a=a_2(\th)$
obtained in \eqref{eq:a12}.
It would also be interesting to study the transition curves $a=a_k(\th)$
for higher instanton corrections $Z_{k\geq3}$, which we will leave for a future problem.

\section{Conclusion and open problems}\label{sec:discussion}
In this paper we have examined the phase diagram of
$q$-deformed Yang-Mills theory on $S^2$ conjectured in \cite{Jafferis:2005jd} and
found numerical evidence for this conjecture using the exact partition
function at finite $N$ \eqref{eq:ZN-det}.
We found that the $1/N$ correction to the instanton contribution, in particular the prefactor
of instanton, is important for the understanding of 
the splitting of phase transition curves at non-zero
$\th$. Our analysis heavily relied on numerics and it is desirable
to find a more analytic method to study the phase diagram.

There are various open problems.
We have seen that the instanton correction in $q$YM 
has an interesting connection to a $q$-deformation of Laguerre
polynomials, in the sense that the 1-instanton term $Z_1=\xi\Tr M$
reduces to the ordinary Laguerre
polynomial in the limit \eqref{eq:un-1inst}. 
More generally, we observed that the characteristic polynomial of $M$
in \eqref{eq:Mmat-q} reduces to that of $\til{M}$ in \eqref{eq:tilM} 
in this limit \eqref{eq:laguerre-det}.
However, at present it is not clear to us how 
the instanton correction in $q$YM 
is related to the standard definition of $q$-Laguerre
polynomials, also known as the generalized Stieltjes-Wigert polynomials (see e.g. \cite{DLMF2}).
Understanding such a relation could be a key towards an analytic proof of
the conjectured form of $M$ in \eqref{eq:Mmat-q}. 
Also, it is important to find the analytic form of the instanton prefactor
$f_k(t,g_s)$ in the $q$-deformed case which plays an important role
for the phase diagram at non-zero $\th$.
We observed numerically that only the anti-symmetric representations 
of 
instantons are relevant and other representations
are suppressed in the large $N$ limit \eqref{eq:r1}, 
as conjectured in \cite{Jafferis:2005jd}. 
It would be interesting to understand the physical origin of this phenomenon.
It would also be important to understand
the implication of this phase structure
for the black hole physics.
According to \cite{Vafa:2004qa,Aganagic:2004js} the coefficient 
$\Om({\vec{m}})$ in \eqref{eq:Zw}
is related to the black hole entropy, and
it would be very interesting to study its large $N$ behavior.
It is argued in 
\cite{Vafa:2004qa,Dijkgraaf:2005bp,Aganagic:2006je} that the non-perturbative
$\mathcal{O}(e^{-N})$ effect is responsible for the failure of chiral factorization of
partition function and it has an interesting consequence in the dual
spacetime picture. It would be very interesting to
study such $\mathcal{O}(e^{-N})$ effects in the $q$-deformed Yang-Mills theory
from the viewpoint of resurgence, in a similar manner as
the GWW model
studied in \cite{Marino:2008ya,Ahmed:2017lhl}.
Some progress in this direction 
for the 2d Yang-Mills theory on a torus will be reported elsewhere \cite{Sakai}.
Also, we expect that the chiral partition function of $q$YM
receives
``membrane instanton corrections'' given by
the Nekrasov-Shatashvili limit of the refined topological string
on $X_p$ from the general argument in \cite{Hatsuda:2013oxa}.
Note that it was shown in \cite{Aganagic:2012si} that the 
refined topological string on $X_p$ is related to a two parameter $(q,t)$-deformation of
2d Yang-Mills theory and the large $N$ phase structure of $(q,t)$-deformed 2d Yang-Mills on $S^2$
was studied in \cite{Kokenyesi:2013nxa}.
It would be very interesting to investigate this direction further.

\vskip8mm
\acknowledgments
I would like to thank Daniel Jafferis
for correspondence.
This work  was supported in part by JSPS KAKENHI Grant No. 16K05316.

 
\end{document}